\documentclass[twocolumn,showpacs,amsmath,amssymb,aps,pre,reprint,longbibliography]{revtex4-2}
\usepackage[pdftex]{graphicx}
\usepackage{amsmath}

\begin{document}

\title{
Ratchet Effects For Swarmalators on Asymmetric Substrates } 
\author{
B. Albert$^{1}$, A. Lib{\' a}l$^{1}$, C. J. O. Reichhardt$^{2}$ and C. Reichhardt$^{2}$, 
} 
\affiliation{
$^{1}$ Mathematics and Computer Science Department, Babe{\c s}-Bolyai University, Cluj 400084, Romania\\
$^{2}$ Theoretical Division and Center for Nonlinear Studies,
Los Alamos National Laboratory, Los Alamos, New Mexico 87545, USA}

\date{\today}
\begin{abstract}

  Swarmalators are particles which have internal degrees of freedom in the form of phase oscillators that can synchronize and change the dynamics of the particles. We explore the active matter disk swarmalator model proposed by Adorj\'{a}ni \cite{Adorjani24} where there is an additional attraction or repulsion determined by whether the phases of particles within a certain radius synchronize or anti-synchronize. This model in the absence of swarmalator rules forms a motility-induced phase-separated (MIPS) state at high activity, but when the swarmalator rules are present, a number of additional states appear including an active gel, labyrinth state, frustrated MIPS, and frustrated fluid states. Here we examine ratchet effects for the Adorj\'{a}ni model coupled to asymmetric substrates. We find that swarmalator interactions can strongly enhance or suppress active matter ratcheting depending on the structural state of the system, and show that the ratchet efficiency depends on the substrate strength, substrate lattice spacing, and the motor force on the particles. The gel state exhibits the weakest ratcheting for small substrate spacing and can even produce a reversed ratchet motion,  but shows the strongest ratcheting for large motor forces. The labyrinth state exhibits the strongest ratcheting over a wide range of parameters since the quasi-one-dimensional labyrinth morphology can match the substrate geometry well.
\end{abstract}
\maketitle

\section{Introduction}

Active matter describes systems with self-motility that can be of run-and-tumble or driven diffusive type \cite{Marchetti13,Bechinger16,teVrugt26}.
Realizations of active matter,
which
may involve individual particles or collections of interacting particles
\cite{Marchetti13,Bechinger16,Gompper20},
have been achieved
in biological systems \cite{DiLeonardo10,Bhattacharjee19,Tan22}, soft matter \cite{Palacci13,Buttinoni13,Gompper20}, robotics
\cite{Wang21,BenZion23}, and social systems \cite{Helbing01,Castellano09}.
One of the most interesting phenomena in collective active matter is
the motility-induced phase separation (MIPS) that emerges
when the activity or persistence length is high enough.
In MIPS, there is phase separation even for purely repulsive
particle-particle interactions into a high-density phase
that coexists with a lower-density gas phase
\cite{Fily12,Redner13,Palacci13,Buttinoni13,Bialke15,Cates15,Paoluzzi22}.

In most active matter systems, the interactions between the particles are fixed as a function of time. In the swarmalator model, first introduced
by O'Keefe {\it et al.}~\cite{OKeefe17},
the particles have an internal degree of freedom in the form of
a phase oscillation, permitting the phase dynamics
to be coupled with the particle dynamics.
The original swarmalator model had an infinite interaction length,
and produced a number of distinct dynamical states.
There have been numerous generalizations of swarmalators \cite{Sar26}
including one-dimensional versions \cite{Okeeffe22}, and swarmalators have
been proposed that have
higher-order interactions \cite{Anwar24,Hu25},
additional disorder \cite{Yoon22,Senthamizhan26},
inertia \cite{OKeeffe26},
chirality \cite{Ceron23},
time delays \cite{Blum24,OKeeffe26a,Lumbu26},
pinning effects \cite{Sar23},
dynamics in higher dimensions \cite{Yadav24},
frustration effects \cite{Lizarraga23,Senthamizhan25},
nonreciprocal interactions \cite{Yu25},
thermal effects \cite{Hong23},
and topological effects \cite{Gou26,Louodop26}.
Aspects of swarmalator behavior are relevant for various
biological and robotics systems \cite{Sar26}, and
there have recently been
experimental realizations of colloidal swarmalators \cite{Heuthe25}.

Adorj\'{a}ni {\it et al.} \cite{Adorjani24} introduced a variation
of the swarmalator model for a system of active
stiff elastic disks that undergo run-and-tumble motion
and have interactions which change over time according to the swarmalator
rules. Unlike the original swarmalator model \cite{OKeefe17},
which had infinite interaction length for the swarmalator rules,
in the Adorj\'{a}ni model there is a finite interaction length
extending only a few particle diameters away to a distance $2r_{\rm sw}$.
In the Adorj\'{a}ni model, the particles
have an internal phase $\phi_i\in[0,2\pi)$
that produces an additional attractive or repulsive force
between neighboring disks $i$ and $j$
based on the internal phase difference
$\delta\phi_{ij}=\phi_i-\phi_j$,
the swarmalator parameter $J$,
and the distance separating the particles.
The dynamics of the internal phase $\phi_i$ is determined by the
parameter $K$ as well as the distance to and
the phase difference with
the neighboring particles.
With the swarmalator rules added to the run-and-tumble dynamics,
the system exhibits five additional states as a function of $J$ and
$K$ beyond the single
non-swarmalator MIPS state: active gel, superlattice MIPS, frustrated MIPS,
labyrinth, and frustrated cluster fluid.

Active matter systems can exhibit directed transport or ratchet effects
when they are coupled to an asymmetric substrate.
Ratchet effects have been extensively studied for
non-active particles on asymmetric substrates
such as a sawtooth potential, and can appear when
periodic driving is applied or
when the substrate is flashed in the presence of thermal
fluctuations \cite{Magnasco93,Rousselet94,Bartussek96,Reimann02}.
The ratchet effect arises from a combination of
spatial symmetry breaking by the substrate and
temporal symmetry breaking by the driving.
Typically, ratchet motion consists of a dc flow of particles
along the easy direction of the substrate,
as has been studied for
colloids \cite{Straube14,Tierno14},
granular matter \cite{Farkas02},
vortices in superconductors \cite{Lee99,Wambaugh99},
cold atom systems \cite{MenneratRobilliard99},
quantum systems \cite{Salger09},
and magnetic skyrmions \cite{Reichhardt15a}.
When collective interactions between the particles occur,
it is possible to observe ratchet reversals and commensuration effects
\cite{Villegas05,deSouzaSilva06a,Dinis07a,Mukhopadhyay18}.
In active matter systems, ratchet effects
can occur in the absence of external driving, since they can be
generated by the combination of spatial symmetry breaking from the
substrate or wall interactions with
breaking of detailed balance by the activity
\cite{Reichhardt17a,OByrne22,Metzger26}.
One of the first observations of active ratchets was
performed with bacteria in asymmetric funnel arrays
\cite{Galajda07}, and since then a variety of different active matter
ratchet geometries have been explored
\cite{Wan08,Tailleur09,Angelani11,Reichhardt13,Berdakin13,Ai14,Nikola16},
including ratchets in active nematics \cite{Schimming24}
or rotating asymmetric gears \cite{DiLeonardo10,Kaiser14,Jerez20,Ai16},
as well as various ratchet reversal effects produced by
particle-particle interactions
\cite{Drocco12,McDermott16}.

An open question is whether ratchet effects can also arise in
swarmalator systems. Here we consider the Adorj\'{a}ni model
\cite{Adorjani24} coupled to a one-dimensional asymmetric sawtooth
potential similar to that used in previous works on active matter ratchets.
We show that this system exhibits a strong ratchet effect in the easy direction
along with a non-monotonic ratchet efficiency as a function of
substrate strength, periodicity, and active motor force.
We also show that the different states that appear as
$K$ and $J$ are varied exhibit different ratchet efficiencies,
with regions of strongly increased or decreased ratchet efficiency
compared to a system in which the swarmalator rules are absent.
We demonstrate that substrates can partially break up the
different cluster states,
and that this correlates with enhanced or decreased ratchet
effects depending on the state.
For example, the active gel
displays the weakest ratchet effects for strong substrates and
small substrate lattice constants compared to the MIPS state,
but has a much stronger ratchet effect for high motor forces.
For smaller substrate lattice spacings,
the labyrinth state shows enhanced ratchet effects
compared to the non-swarmalator system due to the
fact that the quasi-one-dimensional structures in the
labyrinth can align with the substrate,
causing the interactions between rows of particles in adjacent
substrate troughs to be reduced.

\section{Model}

\begin{figure}
\includegraphics[width=\columnwidth]{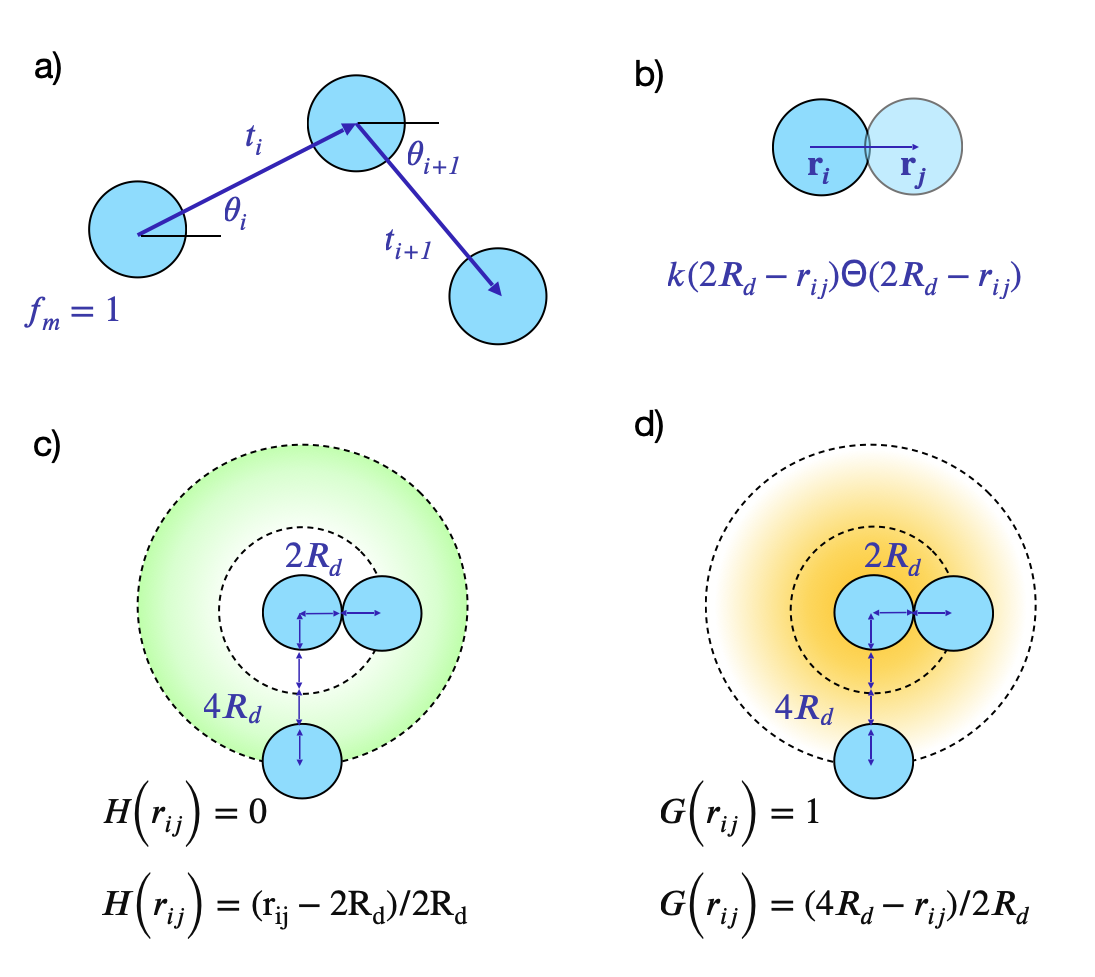}
\caption{Schematics of the active disk swarmalator system.
(a) The basic active disk model showing a disk with a motor force $f_m$
traveling at angle $\theta_i$ and undergoing a tumble to a new angle
$\theta_{i+1}$ after a run time of $t_i$.
(b) The repulsive disk-disk interaction acts out to a distance $R_{d}$.
(c) The additional swarmalator force, which acts over a distance
$r_{\rm sw} < r < 2r_{\rm sw}=4R_d$,
is positive or negative depending on the sign of $J$
and the internal phase difference of the two disks.
(d) The dynamics of the swarmalator phase is determined by disks within
a radius $2r_{\rm sw}=4R_d$, and can be aligning or anti-aligning depending on the
sign of $K$.}
\label{fig:1}
\end{figure}

We consider the Adorj\'{a}ni swarmalator model \cite{Adorjani24} in
a system
with periodic boundary conditions in the $x$ and $y$-directions
that is of length $L = 160$ on a side.
Within the system we place $N = 4000$ run-and-tumble active disks
of radius $R_d$,
corresponding to a density of
$\rho = \pi R^2_{d}/L^2 = 0.49$.
Each disk has an internal phase degree of freedom $\phi_i$.
The run-and-tumble motion is produced by a motor force ${\bf f}_m$
of magnitude $f_m$ 
applied along a running direction $\theta_i \in [0, 2\pi)$.
The running direction is randomly selected after each tumbling interval,
which is set for each disk to  
$t_i \in [\tau, 2\tau]$, where we label $\tau$ the running time.
In Fig.~\ref{fig:1}(a) we show an active disk with a
motor force $f_m$ traveling at an angle $\theta_i$.
After a run time of $t_i$, the particle reorients and moves away in
the new direction $\theta_{i+1}$ for the next run time interval. This
run-and-tumble motion repeats indefinitely.

The disks interact sterically via a short range harmonic repulsive potential
with a radius of $R_{d} = 1.0$.
The disk-disk repulsion is given by
${\bf f}_{\rm dd} = k(2R_d-r_{ij}) \Theta(2R_d-r_{ij}) {\bf \hat{r}}_{ij},$
where $r_{ij}=|{\bf r}_j-{\bf r}_i|$ is the distance between disks $i$ and
$j$, ${\hat {\bf r}}_{ij}=({\bf r}_j-{\bf r}_i)/r_{ij}$ is the unit vector connecting the centers of the disks, and $\Theta$ is the Heaviside step function
which cuts off the interaction for distances
beyond which the disks are no longer in steric contact.
We set the elastic constant to $k = 20.0$.
Figure~\ref{fig:1}(b) illustrates the repulsive interaction between
two disks that approach each other more closely than
$2R_{d}$.

The swarmalator interaction force ${\bf f}_{\rm sw}=f_{\rm sw}{\bf \hat{r}}_{ij}$ can be expressed as:
\begin{equation}
f_{\rm sw} = \sum_j^N J \cos(\delta \phi_{ij})H(r_{ij}) \ ,
\end{equation}
where  $H(r_{ij})=0$ for $r_{ij} \leq r_{\rm sw}$ or $r_{ij}>2r_{\rm sw}$ and $H(r_{ij})=(r_{ij}-r_{\rm sw})/r_{\rm sw}$ for
$r_{\rm sw} < r_{ij} \leq 2r_{\rm sw}$, giving a swarmalator force that linearly increases from zero to a maximum magnitude
as the distance between particles changes
from disk-disk contact to the cutoff distance of $r_{\rm sw}=4R_d$.
Here $\delta\phi_{ij} = \phi_i - \phi_j$ is the phase difference between disks
$i$ and $j$.
For $J<0$, disks with similar phase repel each other and
disks with opposite phase attract each other, whereas for $J>0$, 
disks with similar phases attract and those with
opposite phases repel.
When $J=0$, the swarmalator interaction is turned off
and the model reverts to the normal MIPS behavior found 
for standard active matter particles. 
The operation of the additional swarmalator interaction force,
with a distance dependence given by $H(r_{ij})$ out to a maximum distance
of $4R_d$, is illustrated
in Fig.~\ref{fig:1}(c).

The internal phase $\phi_i$ of the particles is updated using the second swarmalator equation:
\begin{equation}
	\frac{d\phi_i}{dt} = \sum_j^N K \sin(\delta \phi_{ij})G(r_{ij}) 
\end{equation}
This update will align the phases for $K>0$ and anti-align
the phases for $K<0$.
When $K=0$ each particle retains the initial random phase value that was
assigned to it.
The alignment interaction
is strongest for particles that are in
steric contact and linearly decreases to zero at a distance of
$r_{ij} = 2r_{\rm sw}$ giving $G(r_{ij})=1$ for $r_{ij}\leq r_{\rm sw}$, $G(r_{ij})=(2r_{\rm sw}-r_{ij})/r_{\rm sw}$ for $r_{\rm sw} < r_{ij} \leq 2r_{\rm sw}$,
and $G(r_{ij})=0$ for $r_{ij}>2r_{\rm sw}$.
Figure~\ref{fig:1}(d)
illustrates the manner in which
the evolution of the swarmalator
phase $\phi_i$
depends on the inter-particle distance according to the function
$G(r_{ij})$ out to a maximum distance of $4R_d$.

\begin{figure}
\includegraphics[width=\columnwidth]{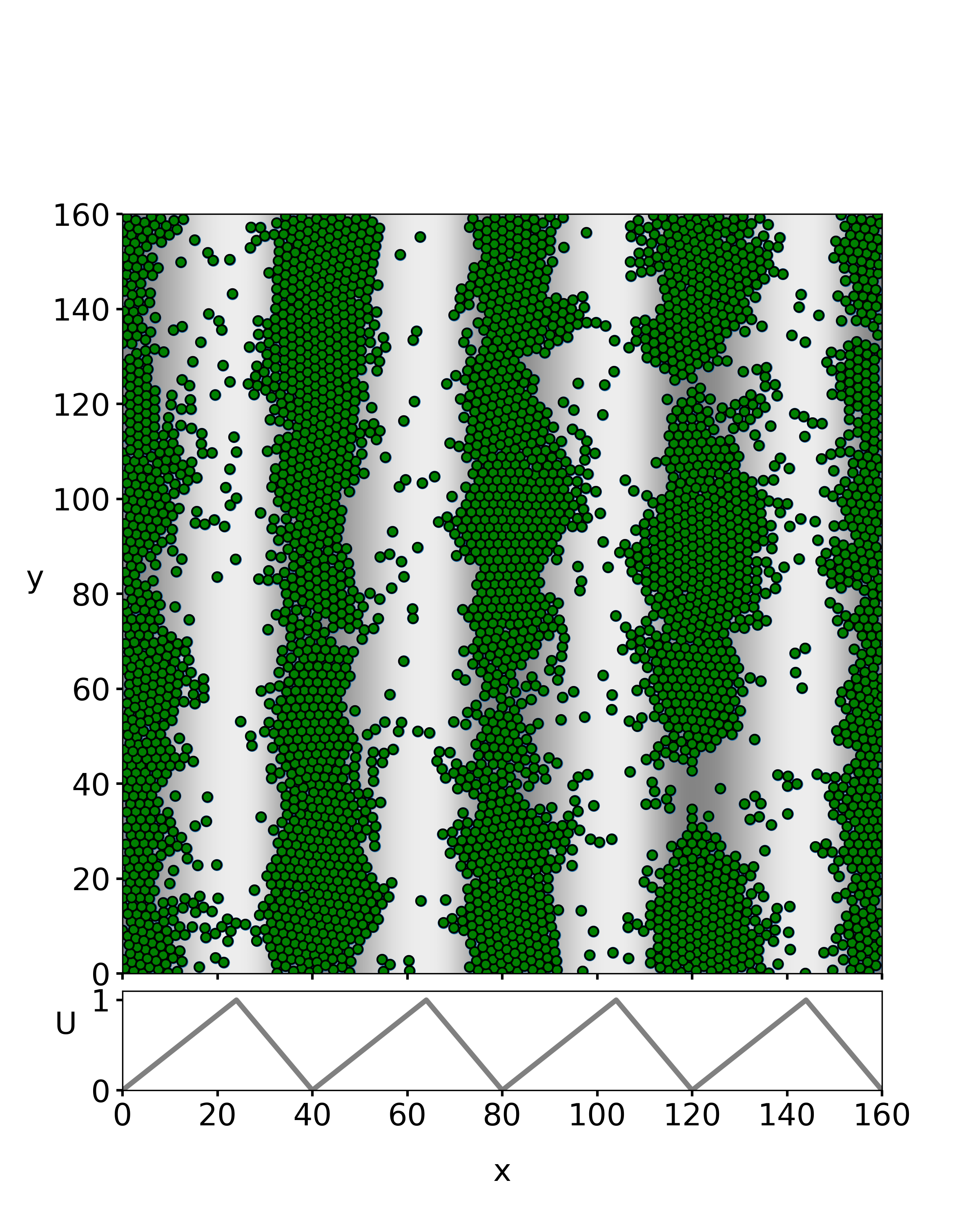}
\caption{Representative image of an asymmetric ratchet potential
with $r_r=0.6$. Green
circles are the particles in the non-swarmalator normal MIPS state
with $J=0$ and $K=0$;
the phase information is suppressed since $J=0$.
The gray shading indicates the depth of the
sawtooth substrate potential.
The lower panel shows a side view of the substrate ratchet potential $U(x)$.
}
\label{fig:3}
\end{figure}

For the substrate force ${\bf f}_r$,
we use a sawtooth
ratchet potential of period $a_r$,
giving a total of $N_r=L/a_r$
periods in our sample.
Each period is split into
a rising fraction $r_r$ giving a rising length of
$d_r=r_r a_r$, and a falling fraction of $1-r_r$ giving
a falling length of $d_f=(1-r_r)a_r$.
The force from the substrate is set
such that the maximum force magnitude exerted
is always $f_{r}^{\rm max}$
in the steeper part of the ratchet tooth, and
the maximum force is correspondingly smaller in the shallower portion
of the ratchet tooth.
For example, for $r_r<0.5$, the force in the
rising portion of the ratchet is $f_r=f_r^{\rm max}$,
while in the falling portion of the ratchet it is
$f_r=-f_r^{\rm max}r_r/(1-r_r)$.
An example of an asymmetrical ratchet with $r_r=0.6$ appears in
Figure~\ref{fig:3},
where the particles clearly accumulate in the low
energy portions of the substrate.

The motion of the particles is obtained
by integrating the following equation of motion:
\begin{equation}
\eta\frac{d{\bf r}_i}{dt} = {\bf f}_{\rm m} + {\bf f}_{\rm dd} + {\bf f}_{\rm sw} +  {\bf f}_{\rm r} \ .
\end{equation}
Here ${\bf f}_{\rm m}$ is the motor force, ${\bf f}_{\rm dd}$ is the disk-disk force, ${\bf f}_{\rm sw}$ is the swarmalator force, and ${\bf f}_{\rm r}$ is the force from the ratchet substrate potential.
We assume overdamped dynamics where we set the damping coefficient to $\eta=1$,
and we use a simulation time step of $dt=0.001$.
To initialize the sample, we place the disks in randomly chosen
non-overlapping positions and set the initial values of
$\theta_i$, ${\phi}_i$, and $t_i$ to random values
chosen with equal probability from the allowed range of each quantity,
following the parameters studied in Ref.~\cite{Adorjani24}.
We measure the average velocity per particle in the $x$ direction,
given by
$\langle V_x \rangle=N^{-1}\sum^N_i {\bf v}_i \cdot {\bf \hat x}$,
where ${\bf v}_i$ is the velocity of particle $i$.
We allow the system to run for $2 \times 10^6$ simulation time steps
before collecting data to avoid transient effects,
and average the velocity over $4 \times 10^6$ steps
as well as over 10 realizations with different random seeds.

\begin{figure}
\includegraphics[width=\columnwidth]{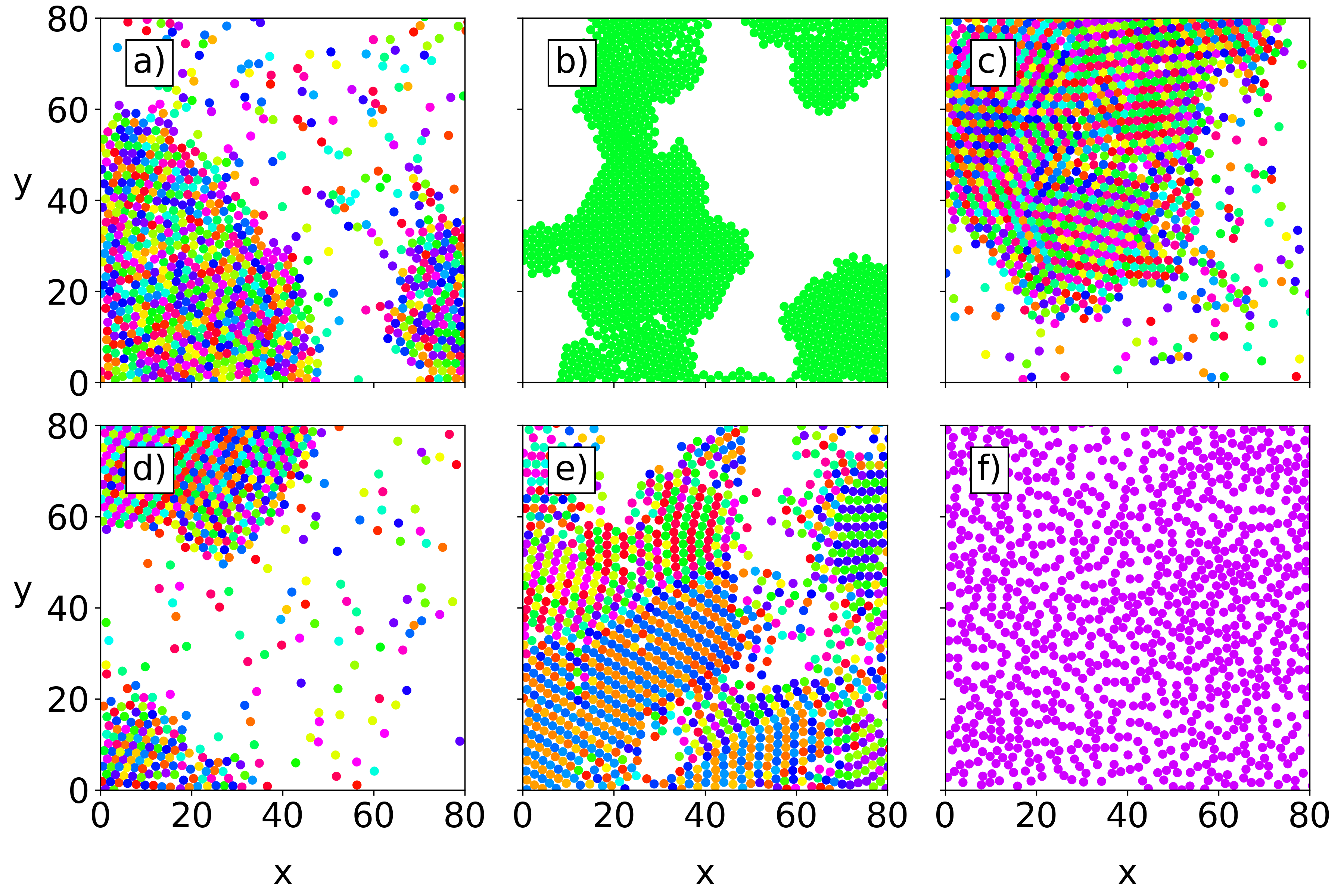}
\caption{Particle locations in the different states of the system that
appear as $J$ and $K$ are varied in a system with no substrate
potential for a density of $\rho=0.5$,
a run time of $\tau=10^5$, and a motor force of $f_m=1.0$.
Color indicates the value of
the phase $\phi_i$ for
each particle, and we show only a quarter of the system for clarity.
(a) Normal MIPS at $J = 0.0$ and $K = 0.0$.
(b) The active gel at $J = 1.0$ and $K = 1.0$.
(c) Superlattice MIPS for $J = 1.0$ and $K = -0.1$.
(d) Frustrated cluster fluid at $J = 1.5$ and $K = -0.1$.
(e) Frustrated MIPS state with $J = -1.5$ and $K = -0.1$.
(f) Labyrinth state at $J = -1.0$ and $K = 0.1$.
}
\label{fig:2}
\end{figure}

When the swarmalator rules are added to the run-and-tumble dynamics,
the system exhibits five additional states as a function of $J$ and
$K$ beyond the single
non-swarmalator MIPS state.
Figure~\ref{fig:2} shows steady-state particle configurations
in these different states in the absence of a substrate
at a density of $\rho=0.5$ and a run time of $\tau=10^5$,
where the color indicates the internal
phase $\phi_i$ of the particle.
When $J = 0$, the swarmalator interactions are turned off and we have a simple
active disk system that 
forms a MIPS state for large activity. This is shown in
Fig.~\ref{fig:2}(a) for a system where we also set $K=0$ so that
the phases of the particles are random and do not change
with time.
For $J=0$ and $K > 0.0001$, the system organizes to a
MIPS state in which
all of the particles have the same phase.
When $J$ and $K$ both have nonzero positive
values, the swarmalator interactions produce
an additional attractive force between particles.
This produces a state that we call an active gel,
shown in Fig.~\ref{fig:2}(b) for $J = 1.0$ and $K = 1.0$.
For positive $J$ and negative $K$,
we obtain a superlattice MIPS at smaller $J$ values,
as shown in Fig.~\ref{fig:2}(c)
for $J = 1.0$ and $K =-0.1$,
and a frustrated cluster fluid for larger values of $J$,
as shown in Fig.~\ref{fig:2}(d) for $J = 1.5$ and $K =-0.1$.
For
repulsive interactions $J < -0.5$
and an anti-phase synchronization value of 
$K < -0.0001$,
we find a stripe-ordered frustrated MIPS state
as shown in Fig.~\ref{fig:2}(e) at $J = -1.5$ and $K = -0.1$.
For $J < -0.5$ and positive $K$, all of the particles organize into the
same phase but the particle-particle repulsion causes
the formation of a labyrinth state or quasi-one-dimensional
chains as illustrated for $J=-1.0$ and $K=0.1$
in Fig.~\ref{fig:2}(f).
Adorj\'{a}ni {\it et al.}~\cite{Adorjani24}
also found some sub-regimes such as an arrested gel state at
$J>0$ and $K=0.0$ where the particle interactions are attractive but
the particles do not change internal phase in response to their neighbors.

\section{Results}

We first consider particles with a run time of
$\tau = 1.5\times 10^5$ in
the six different states highlighted in Fig.~\ref{fig:2},
and vary the ratchet rising fraction $r_r$.
For $r_r = 0.5$, the substrate is symmetric and there is no ratchet effect.
For $r_r > 0.5$, the easy direction for the ratchet effect is in the
positive $x$ direction, while for $r_r < 0.5$, the easy flow is
in the negative $x$ direction.

\begin{figure}
\includegraphics[width=\columnwidth]{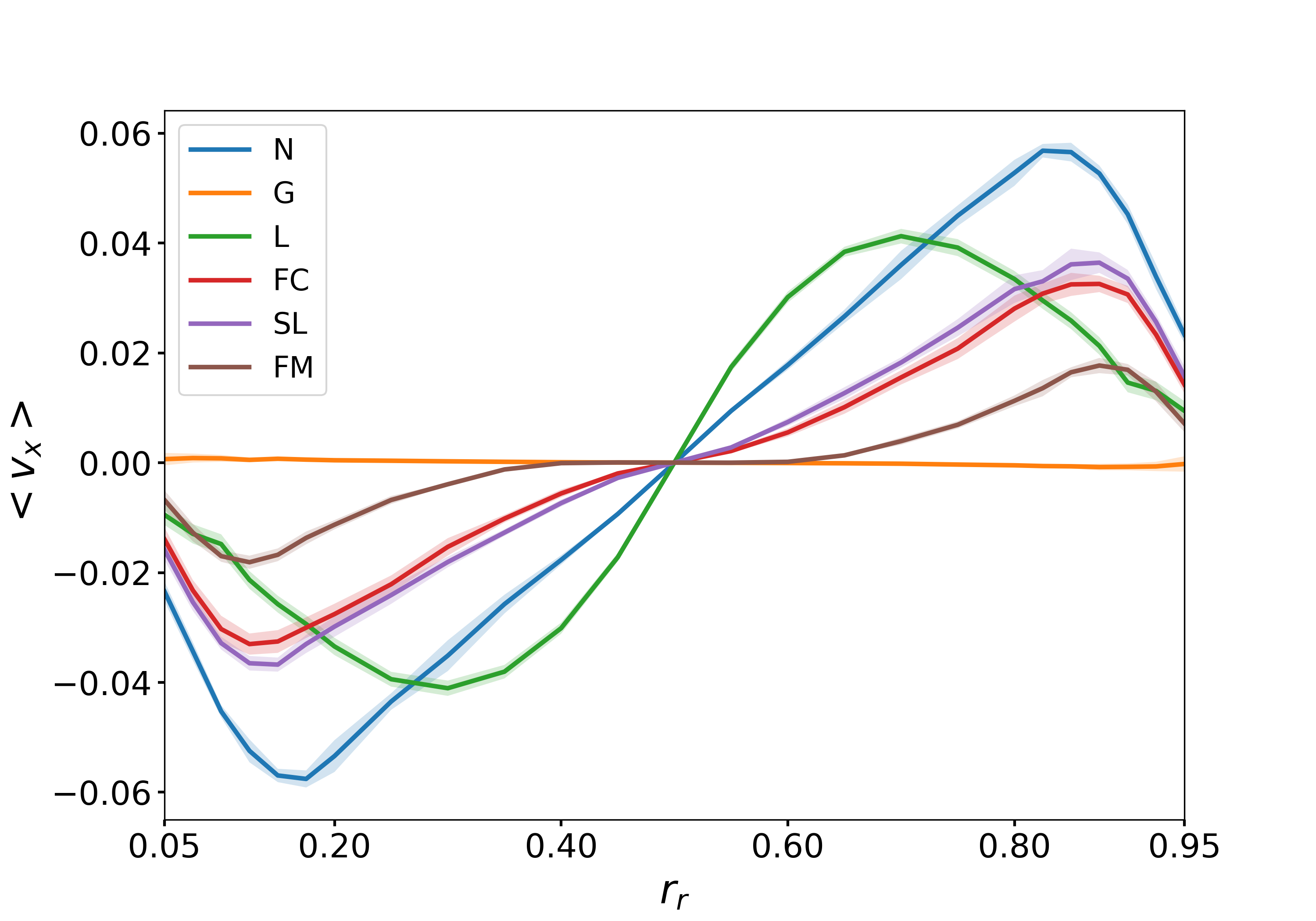}
\caption{ $\langle V_x\rangle$ vs the rising fraction $r_r$
for a system with $a_r=40.0$, $f_r^{\rm max}=1.0$, and $f_m=1.0$ for
the normal MIPS
with $J=0$ and $K=0$ (N, blue),
active gel with $J=1.0$ and $K=1.0$ (G, orange),
labyrinth with $J=-1.0$ and $K=0.1$ (L, green),
frustrated cluster fluid with $J=1.5$ and $K=-0.1$ (FC, red),
superlattice MIPS with $J=1.0$ and $K=-0.1$ (SL, maroon), and
frustrated MIPS with $J=-1.5$ and $K=-0.1$ (FM, brown).
At $r_r = 0.5$ the substrate is symmetric and there is no ratchet effect.
For $r_{r} > 0.5$ the easy flow direction is along
$+x$ and for $r_r < 0.5$ the easy flow direction is along $-x$.}
\label{fig:4}
\end{figure}

In Fig.~\ref{fig:4} we plot the average velocity per particle
$\langle V_x\rangle$ versus the ratchet rising fraction $r_r$
for the different dynamical regimes highlighted in
Fig.~\ref{fig:2}. The
active gel state shows almost no ratchet effect
or a slightly negative ratchet effect,
while all of the other states exhibit a strong ratchet effect in the
easy flow direction.
The normal MIPS state has the overall strongest ratchet effect
near $r_r = 0.8$, where $\langle V_x\rangle = 0.058$.
The ratchet effect is reduced in the
superlattice, frustrated cluster fluid, and frustrated MIPS
states.
When $\langle V_x\rangle \approx 0.06$, the maximum value we
observe, the total distance traveled
on average by a particle during our averaging time of $n_t=4\times 10^6$
steps is
$d_{\rm tot} = \langle V_x\rangle n_t dt = 240$,
equivalent to six lattice periods of the substrate.
For the frustrated MIPS and cluster fluid states, the
ratcheting is less effective and particles move at most
a distance of two to three substrate lattice
constants during $n_t$ simulation
time steps.
In the ranges $0.3 < r_r <0.5$ and $0.5 < r_r < 0.7$,
the labyrinth state ratchets more effectively than all other states,
including the normal MIPS states, but outside of these windows the
ratchet efficiency drops off more quickly for the labyrinth state than
for all other states.

\begin{figure}
\includegraphics[width=\columnwidth]{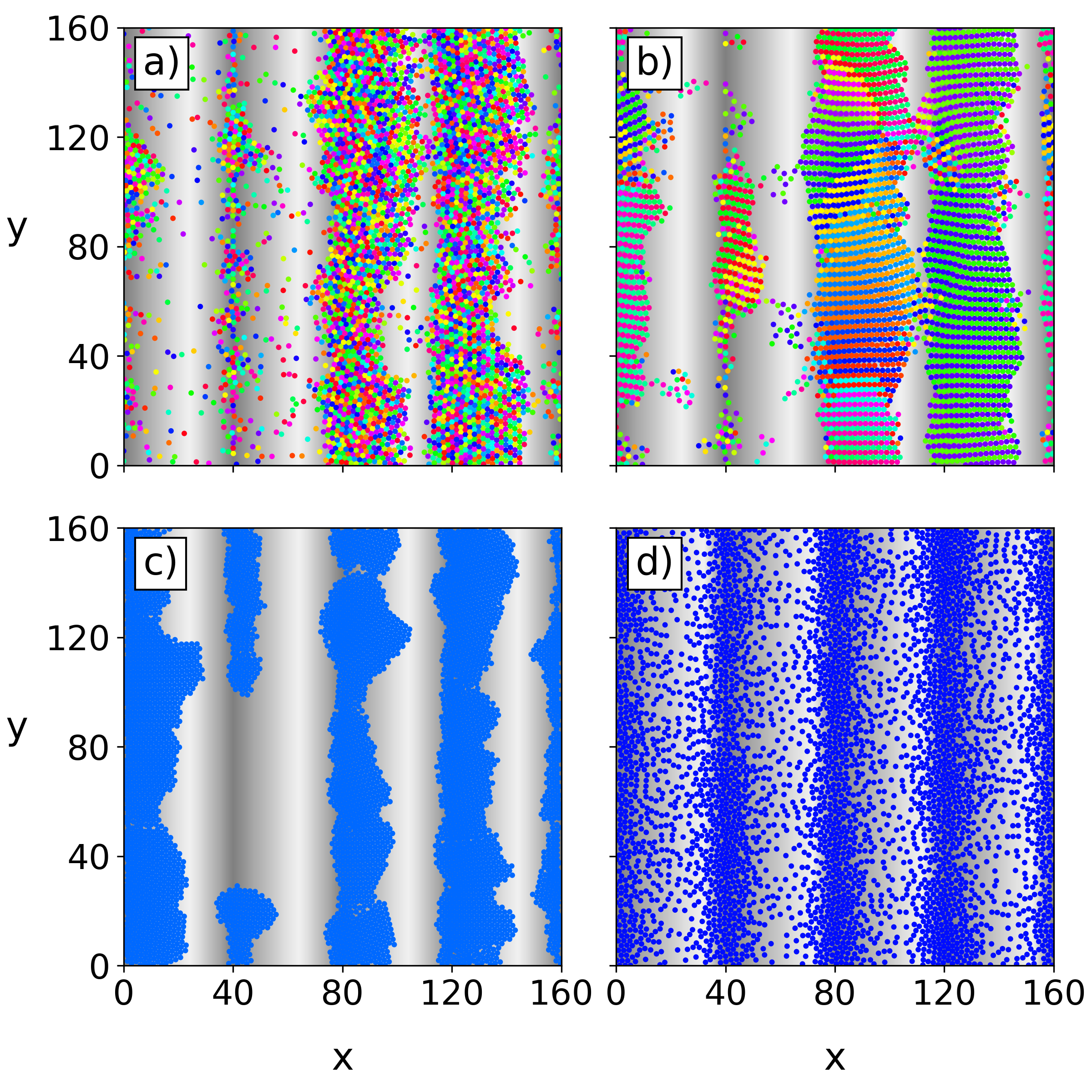}
\caption{Snapshots of the particle positions for the system in
Fig.~\ref{fig:4} with $a_r=40.0$, $f_r^{\rm max}=1.0$, and $f_m=1.0$.
Colors indicate the phase of each particle.
(a) The normal MIPS state with $J=0$ and $K=0$
at $r_r = 0.8$, close to the optimal ratchet effect.
(b) The frustrated MIPS state with $J=-1.5$ and $K=-0.1$ at $r_r = 0.8$,
showing cluster formation.
(c) The active gel state with $J=1.0$ and $K=1.0$
at $r_r = 0.8$, where there is no ratchet effect.
(d) The labyrinth state with $J=-1.0$ and $K=0.1$ at $r_r = 0.6$,
which is close to the optimal $r_r$ for this state.}
\label{fig:5}
\end{figure}

In Fig.~\ref{fig:5} we illustrate the particle positions and phases
for the system in Fig.~\ref{fig:4}.
Figure~\ref{fig:5}(a) shows the optimal ratchet conditions of
$r_r = 0.8$ for the normal MIPS state.
Large portions of the particles are concentrated in
a pair of clusters in neighboring
substrate minima.
When one cluster fills up a substrate minimum,
some individual particles on the edges of the cluster can easily jump to
an adjacent minimum,
while the more sparsely occupied substrate minima are able to better
align the particles with the direction of the substrate troughs.
In Fig.~\ref{fig:5}(b) we show the frustrated MIPS state at
$r_r = 0.8$ where the ratchet efficiency is lower than
in the normal MIPS state.
In this case, although the clusters themselves are aligned along the $y$
direction of the substrate trough, the stripes of frustrated phase orientation
are perpendicular to the substrate troughs and run along the $x$
direction.
In the normal MIPS state, once a particle escapes a cluster
it can move freely as an individual particle
and jump over the barrier separating adjacent
substrate minima.
For the frustrated MIPS, superlattice and frustrated cluster systems,
there are fewer instances of individual particles escaping the cluster.
This is because the swarmalator interaction introduces a
longer range attractive force between particles,
which tends to create smaller clusters of particles
that are not as mobile as
individual particles and therefore
have a lower probability of jumping over the maximum force barrier
in the easy flow direction.
As a result, the ratchet efficiency for these states is reduced.
An example of this clustering
appears in Fig.~\ref{fig:5}(b)
where there are some smaller clusters just to the left of the center of
the system that
move more slowly than the running speed $f_m$ of
individual particles.
In contrast, in the normal MIPS state with $J=0$
the particles experience only the short range repulsive
disk-disk interactions and
are not subjected to attractive swarmalator forces.
Figure~\ref{fig:5}(c) illustrates the active gel at $r_r = 0.8$.
Here all of the particles experience a
longer range swarmalator attractive interaction force,
so in general particles remain trapped in the gel
and do not escape it.
The gel itself can slowly move around,
but the net force along the $x$ direction on a
large cluster is generally much weaker than the
barrier force, so the ratchet effect is strongly reduced.
The slight negative ratchet effect in the gel occurs
when the gel forms a large cluster that is centered in the
substrate minimum but does not fit in the minimum, so that a portion of
the cluster ends up extending over the barrier in the
hard direction since the minimum is less wide in this direction,
and some particles spill into the adjacent substrate trough. 
As a result the particles in this portion of the gel
feel a stronger substrate force along the hard direction, and effectively
slowly pull the entire gel in the hard direction.
In general, increased attraction between particles
reduces the ratchet effect since it increases the size of the
clusters, and larger clusters move more slowly.
This can also be viewed as a reduction in the effective activity,
which would then reduce the active ratchet motion \cite{Reichhardt17a}.

Figure~\ref{fig:5}(d) shows an image of the labyrinth state
at $r_r = 0.6$ where the efficiency of the ratchet is higher 
than in the other five states.
Compared to the other images in Fig.~\ref{fig:5}, the morphology of the
labyrinth state is quite distinctive.
Overall the system has a more fluid character, and contains
quasi-one-dimensional chains
of particles that extend away from the substrate minima along
the $x$ direction.
The existence of these chains
allow particles to more easily escape into the next substrate minimum
along the easy flow direction and produce
an enhanced ratchet effect.
At this same value of $r_r=0.6$, the MIPS state is more compact
than the configuration shown in Fig.~\ref{fig:5}(a) and contains
fewer single particles and small clusters.
For higher $r_r>0.6$, the particles in the labyrinth state
become more localized within the substrate minima
and the chains along the $x$ direction no longer form,
reducing the amount of hopping between substrate minima that can
occur.
An interesting feature of the labyrinth phase is that
strings of particles form even in the absence of a substrate
despite the fact that there is no attractive term in the interaction
potential.
In the labyrinth state, the swarmalator forces
introduce an additional repulsion term
so that the hard disks have a central repulsive core surrounded by
a longer range softer step repulsion.
It is known from work on nonactive systems with
similar two-step or core-corona
repulsive potentials that the particles can form
stripe-like patterns when the
particle density is sufficiently large \cite{Malescio03,Glaser07}.
When the particles overlap at larger distances on the soft part of the
repulsive potential, there is a weak energy cost for moving two
particles closer together, so that the lowest energy state would be to have
no overlap of the particles at all. If the particle density is too high
for this to be possible, however, the system can reduce its energy
by rearranging such that the particles move closer together along one
direction in order to eliminate overlap along the perpendicular direction.
Such an energy reduction is only possible if the outer or corona portion of the
interaction potential is sufficiently soft that changes in the spacing between
the particles do not change the interaction energy very much.
As shown in Fig.~\ref{fig:5}(d), when a system of this type is placed on
an asymmetric substrate, the overlaps of the particles in the low
energy region of the substrate can be reduced by having some particles
form chains that extend along the positive $x$ direction.
These chains incur some additional energy cost from the substrate potential,
but if the reduction in energy achieved by reducing the particle overlap is
sufficiently high, chain formation is favored.
Particles at the ends of the chains have an increased probability of being
able to hop successfully into the next substrate trough.
In particular, when the run direction of a particle at the end of
the chain is aligned away from the chain and
along the easy substrate direction,
there is no energy cost for the particle to leave the chain.
As $r_r$ increases above $r_r=0.6$,
the energy barrier from the substrate becomes
high enough that chains are no longer
able to align with the $x$ direction
and the ratchet efficiency of the labyrinth state drops.

\begin{figure}
\includegraphics[width=\columnwidth]{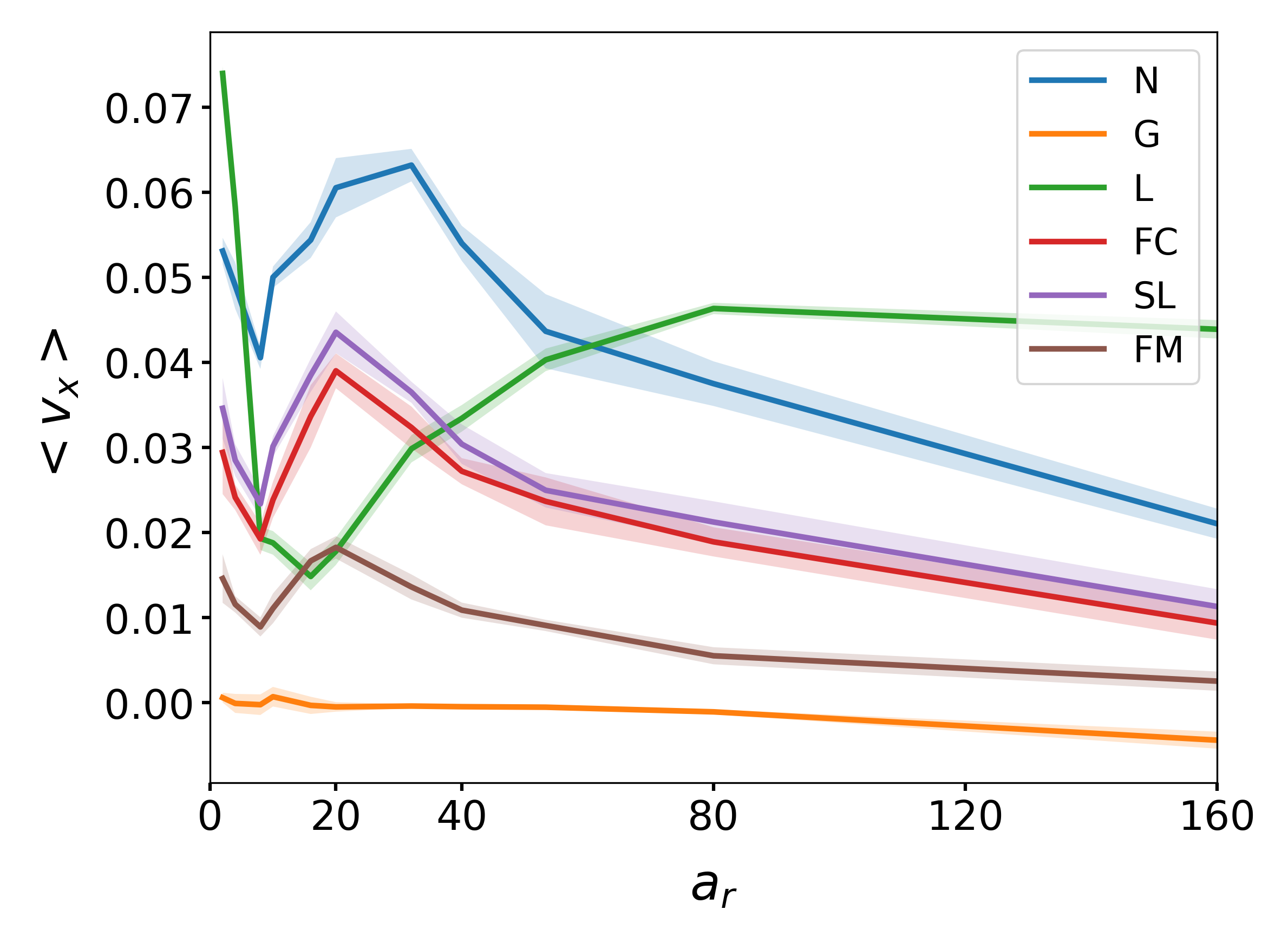}
\caption{$\langle V_x\rangle$ vs the substrate lattice spacing
$a_r$ in a system with $r_r=0.8$, $f_r^{\rm max}=1.0$, and $f_m=1.0$ for
the normal MIPS with $J=0$ and $K=0$ (N, blue),
active gel with $J=1.0$ and $K=1.0$ (G, orange),
labyrinth with $J=-1.0$ and $K=0.1$ (L, green),
frustrated cluster fluid with $J=1.5$ and $K=-0.1$ (FC, red),
superlattice MIPS with $J=1.0$ and $K=-0.1$ (SL, maroon),
and frustrated MIPS with $J=-1.5$ and $K=-0.1$ (FM, brown).
}
\label{fig:6}
\end{figure}

In Fig.~\ref{fig:6} we plot
$\langle V_x\rangle$ versus the substrate lattice spacing $a_r$
for the same states as in Fig.~\ref{fig:4}
at a rising fraction of $r_r = 0.8$
for $a_r = 160$, 80, 53.33, 40, 20, 16, 10, 8, 4,
and 2, corresponding to a number of substrate minima of
$N_r = 1$, 2, 3, 4, 8, 10, 16, 20, 40, and 80, respectively.
For small $a_r < 10$, the labyrinth state
exhibits the largest ratchet effect followed by the
normal MIPS and frustrated MIPS states,
while the active gel state has no ratchet motion.
For $10 < a_r < 60$, the normal MIPS state has the most effective ratchet
motion, while for $a_r > 60$ the labyrinth state once again has the
most efficient ratchet motion.
In general, the states in order of decreasing ratchet efficiency are:
superlattice MIPS, frustrated cluster fluid, frustrated MIPS, and
the active gel. At large $a_r$ the
active gel state has a slightly negative ratchet motion.
The normal MIPS, frustrated cluster fluid, superlattice MIPS,
and frustrated MIPS states all have a local maximum in the
ratchet efficiency near $a_r = 20$.
The structure in these four phases is variations on the MIPS clusters, and
for fixed particle density the size of the clusters in each case is roughly
the same. This cluster size
matches best to the substrate lattice spacing at $a_r = 20$.

\begin{figure}
\includegraphics[width=\columnwidth]{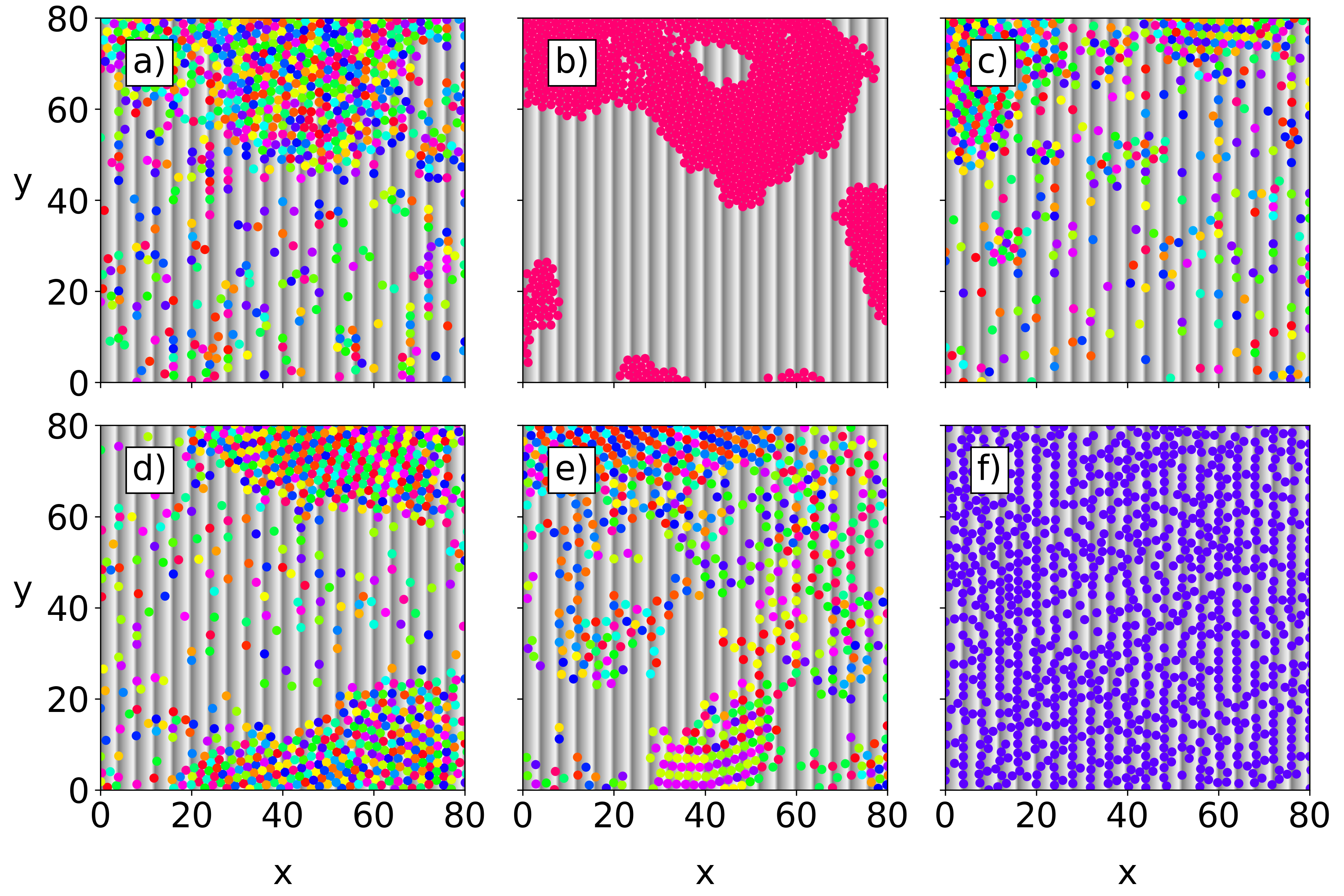}
\caption{Particle locations for the system in
Fig.~\ref{fig:6} with $r_r=0.8$, $f_r^{\rm max}=1.0$, and $f_m=1.0$
at $a_r = 4.0$. Color indicates the value of
the phase $\phi_i$ for each particle,
and we show only a quarter of the system for clarity.
(a) Normal MIPS at $J = 0.0$ and $K = 0.0$.
(b) The active gel at $J = 1.0$ and $K = 1.0$.
(c) Superlattice MIPS for $J = 1.0$ and $K = -0.1$.
(d) Frustrated cluster fluid at $J = 1.5$ and $K = -0.1$.
(e) Frustrated MIPS state with $J = -1.5$ and $K = -0.1$.
(f) Labyrinth state at $J = -1.0$ and $K = 0.1$.
The labyrinth state shows a strong ratchet effect while the
active gel state shows no ratchet effect.
}
\label{fig:7}
\end{figure}

In Fig~\ref{fig:7} we illustrate the particle positions
for the different states from the
system in Fig.~\ref{fig:6} with $r_4=0.8$ at $a_r = 4.0$.
The normal MIPS state appears in Fig.~\ref{fig:7}(a), while
Fig.~\ref{fig:7}(c,d,e) shows the superlattice MIPS,
frustrated cluster fluid, and frustrated MIPS states,
respectively.
In all four of these states,
MIPS clusters appear that are relatively
circular and that therefore
do not align with the substrate, resulting in a
weaker coupling to the substrate and hence a weaker ratchet effect.
In the normal MIPS state in Fig.~\ref{fig:7}(a), isolated patches of particles
are able to achieve
some quasi-one-dimensional substrate alignment, and as a result
the ratchet effect is more efficient
for the normal MIPS state than for the MIPS-like phases
with swarmalator forces.
The active gel state in Fig.~\ref{fig:7}(b) is very weakly
coupled to the substrate and has very little visible evidence of the presence
of a substrate; as a result, the ratchet effect is absent.
The labyrinth state, illustrated in Fig.~\ref{fig:7}(f),
has a stronger ratchet effect than the other states.
The quasi-one-dimensional chains of the labyrinth state
can easily align in the $y$ direction to match the substrate at
smaller values of $a_r<10$.
This permits stronger coupling to the substrate,
leading to the enhanced ratchet effect.

\begin{figure}
\includegraphics[width=\columnwidth]{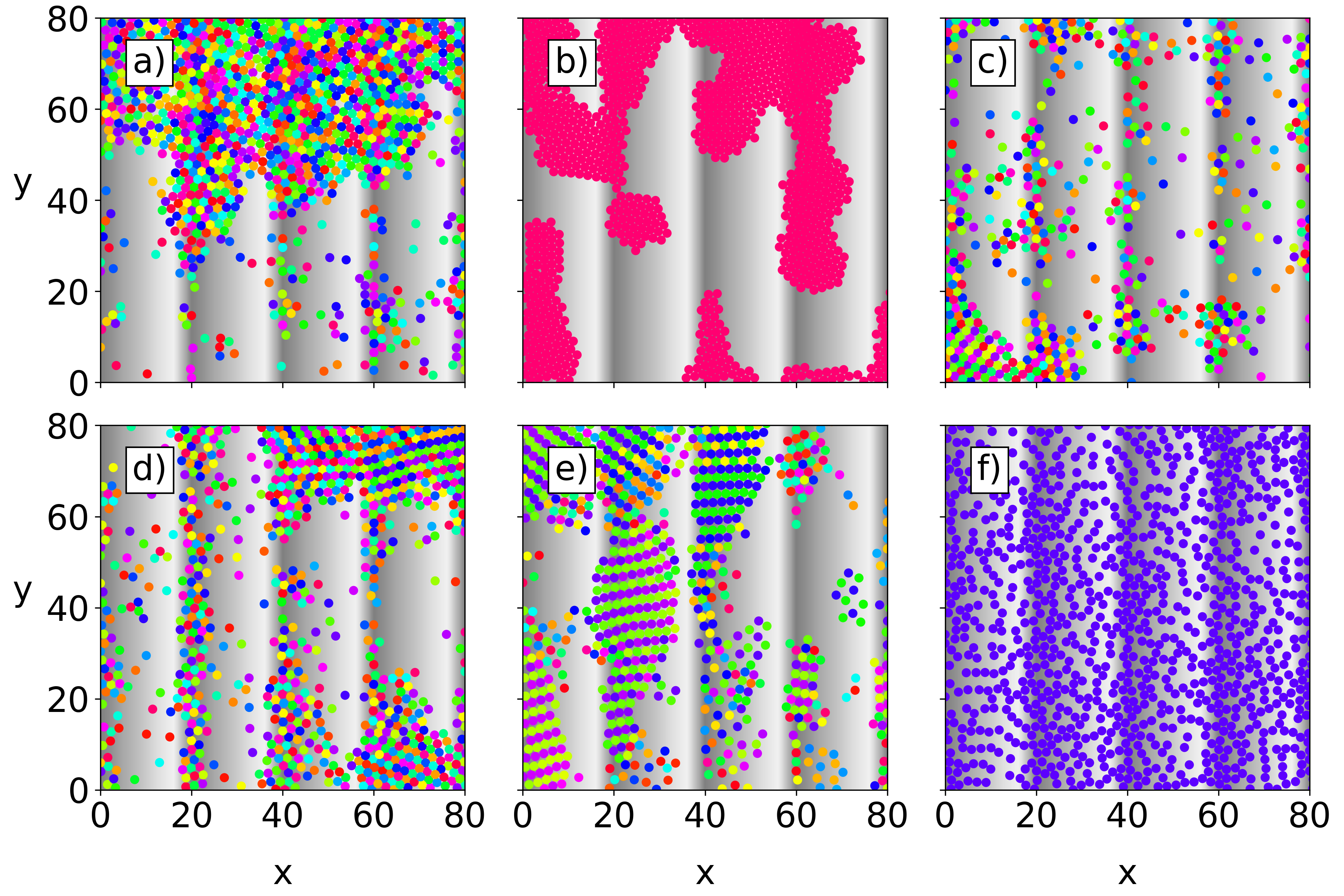}
\caption{Particle locations for the system in Fig.~\ref{fig:6}
with $r_r=0.8$, $f_r^{\rm max}=1.0$, and $f_m=1.0$
at
$a_r = 20.0$.
Color indicates the value of the phase $\phi_i$ for each particle,
and we show only a quarter of the system for clarity.
(a) Normal MIPS at $J = 0.0$ and $K = 0.0$.
(b) The active gel at $J = 1.0$ and $K = 1.0$.
(c) Superlattice MIPS for $J = 1.0$ and $K = -0.1$.
(d) Frustrated cluster fluid at $J = 1.5$ and $K = -0.1$.
(e) Frustrated MIPS state with $J = -1.5$ and $K = -0.1$.
(f) Labyrinth state at $J = -1.0$ and $K = 0.1$.
The labyrinth state has a reduced ratchet effect
while the MIPS states have a strong ratchet effect.}
\label{fig:8}
\end{figure}

For $10 < a_r < 40$, the larger lattice spacing makes it possible for the
MIPS-like states to align with the substrate along the $y$ direction.
This increases the coupling with the substrate, causing a corresponding
increase in the ratchet effect. 
In Fig.~\ref{fig:8}(a,b,c), we show the particle
configurations at $a_r=20.0$ and $r_r=0.8$
for the normal MIPS, active gel and superlattice MIPS states,
respectively,
while Fig.~\ref{fig:8}(d,e,f) illustrates the configurations in
the frustrated cluster fluid, frustrated MIPS, and labyrinth states,
respectively.
Here the MIPS-like states show stronger alignment
with the $y$ direction compared to the
smaller $a_r$ system in Fig.~\ref{fig:7},
while the labyrinth state has weaker alignment in $y$ because the wider
substrate troughs do not match the narrow quasi-one-dimensional chains that
form in this state.

\begin{figure}
\includegraphics[width=\columnwidth]{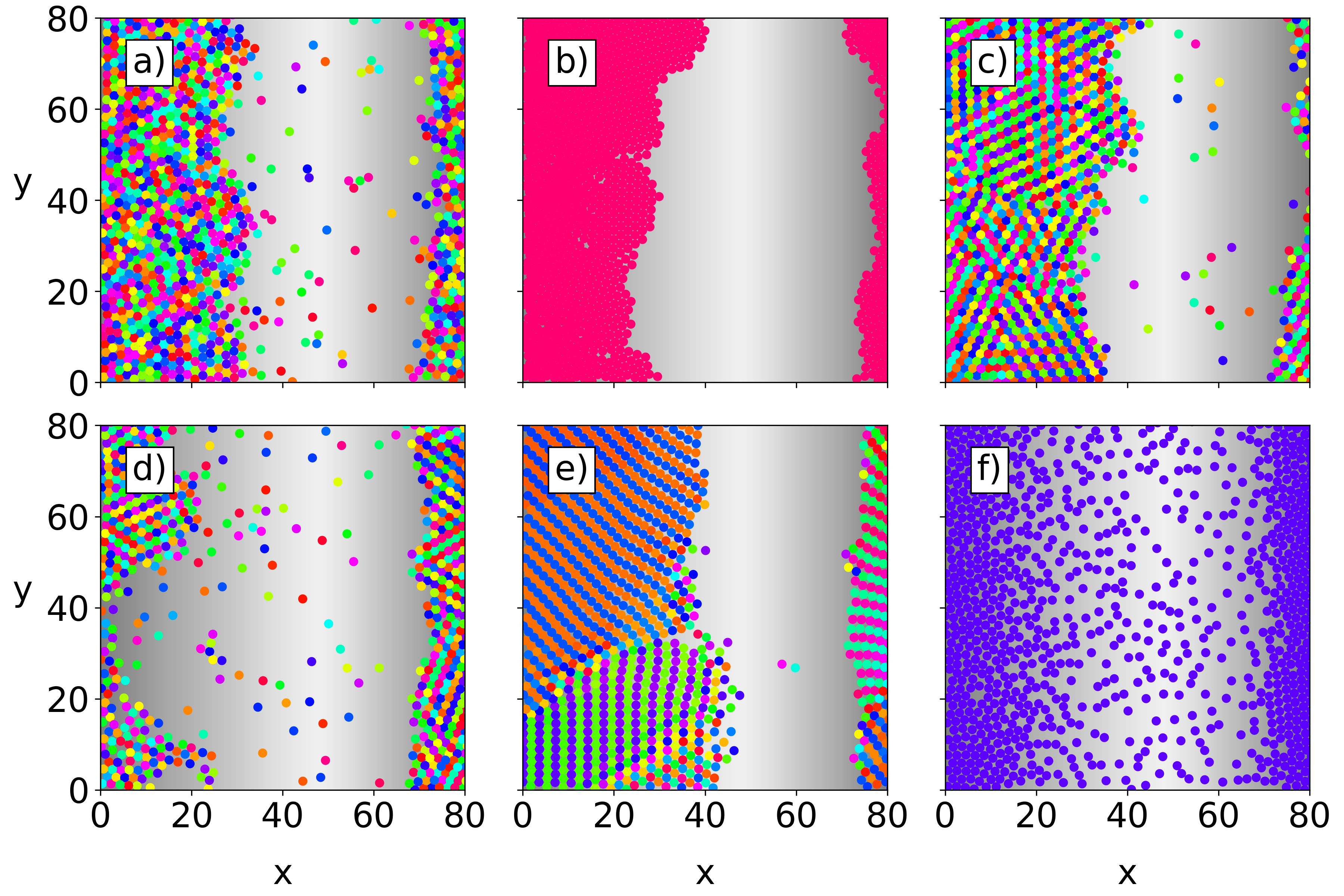}
\caption{Particle locations for the system in
Fig.~\ref{fig:6} with $r_r=0.8$, $f_r^{\rm max}=1.0$, and $f_m=1.0$
at $a_r = 80.0$.
Color indicates the value of the phase $\phi_i$ for each particle,
and we show only a quarter of the system for clarity.
(a) Normal MIPS at $J = 0.0$ and $K = 0.0$.
(b) The active gel at $J = 1.0$ and $K = 1.0$.
(c) Superlattice MIPS for $J = 1.0$ and $K = -0.1$.
(d) Frustrated cluster fluid at $J = 1.5$ and $K = -0.1$.
(e) Frustrated MIPS state with $J = -1.5$ and $K = -0.1$.
(f) Labyrinth state at $J = -1.0$ and $K = 0.1$.
The labyrinth state has the strongest ratchet effect
while the MIPS states have a reduced ratchet effect.
}
\label{fig:9}
\end{figure}

In Fig.~\ref{fig:9} we show the particle configurations
for the system in Fig.~\ref{fig:6} with $r_r=0.8$ at $a_r = 80$.
The labyrinth state in Fig.~\ref{fig:9}(f)
has the strongest ratchet effect.
The normal MIPS, frustrated MIPS, superlattice MIPS, and
frustrated cluster fluid in Fig.~\ref{fig:9}(a,e,c,d)
have a reduced ratchet effect,
while the active gel in Fig.~\ref{fig:9}(b)
shows a weak negative ratchet effect.
In the labyrinth state, the system is more
fluid-like and the particles extend out into the
higher energy portions of the substrate,
while the other states are much more compact.
In general, the ratchet effects are reduced for
these large lattice periods,
since the particles must move a large distance in order to jump into
the neighboring substrate minimum.

\begin{figure}
\includegraphics[width=\columnwidth]{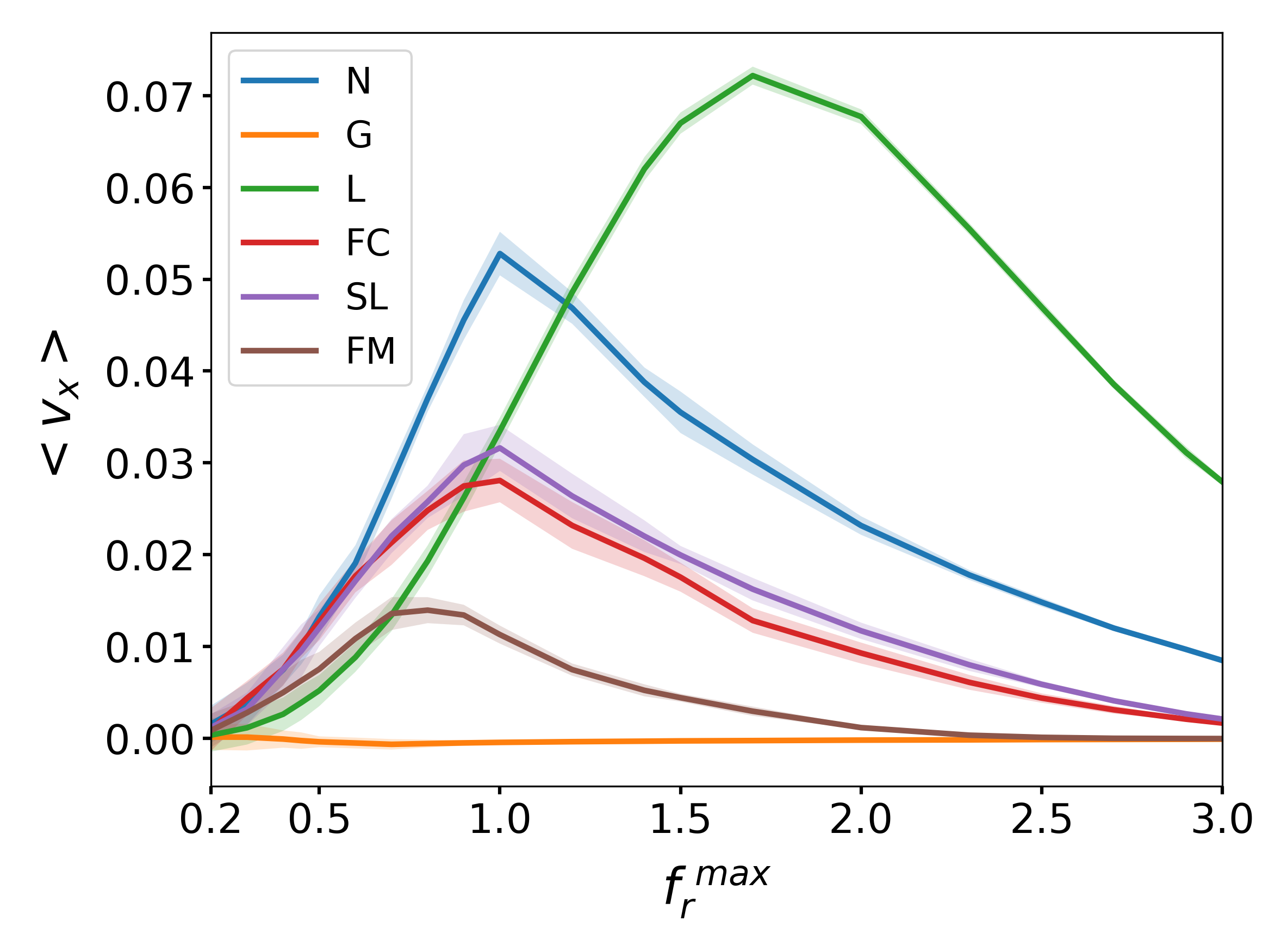}
\caption{$\langle V_{x}\rangle$ vs the maximum substrate force $f_r^{\rm max}$
for a system with $r_r = 0.8$, $a_r = 40$, and $f_m=1.0$
for the normal MIPS with $J=0$ and $K=0$ (N, blue),
active gel with $J=1.0$ and $K=1.0$ (G, orange),
labyrinth with $J=-1.0$ and $K=0.1$ (L, green),
frustrated cluster fluid with $J=1.5$ and $K=-0.1$ (FC, red),
superlattice MIPS with $J=1.0$ and $K=-0.1$ (SL, maroon),
and frustrated MIPS with $J=-1.5$ and $K=-0.1$ (FM, brown).
At $f_r^{\rm max} = 0.0$ there is no ratchet effect and the
MIPS phases show an optimal ratchet effect
when the maximum substrate force matches
the maximum motor force, $f_r^{\rm max}=f_m=1.0$.
The maximum ratchet efficiency of 
the labyrinth state falls at a higher value of $f_{r}^{\rm max}$.}
\label{fig:10}
\end{figure}

Figure~\ref{fig:10} shows $\langle V_{x}\rangle$
versus the maximum substrate strength $f_{r}^{\rm max}$
for a system with $r_r = 0.8$ and $a_{r} = 40$.
For $f_{r}^{\rm max} = 0.0$ the symmetry breaking is lost and
there is no ratchet effect in any of the
states, while for large $f_{r}^{\rm max}$ the particles become permanently
confined by the substrate minima, are unable to hop between minima,
and the ratchet effect is destroyed.
At $f_{r}^{\rm max} = 0.9$, the normal MIPS state has the greatest
ratchet efficiency
followed by the superlattice MIPS, frustrated cluster fluid,
frustrated MIPS, and labyrinth states.
The active gel has almost no ratchet effect.
The peak in the ratchet efficiency near $f_{r}^{\rm max} = 1.0$
that appears for the different MIPS states
also coincides with the value of the motor force,
$f_m=1.0$.
The labyrinth state generally has weak ratcheting
for lower $f_{r}^{\rm max}$ but develops an enhanced ratchet effect
for $f_{r}^{\rm max} > 1.25$,
with a peak in the ratcheting motion appearing near
$f_{r}^{\rm max} = 1.75$ followed by a decrease
at higher $f_{r}^{\rm max}$.

\begin{figure}
\includegraphics[width=\columnwidth]{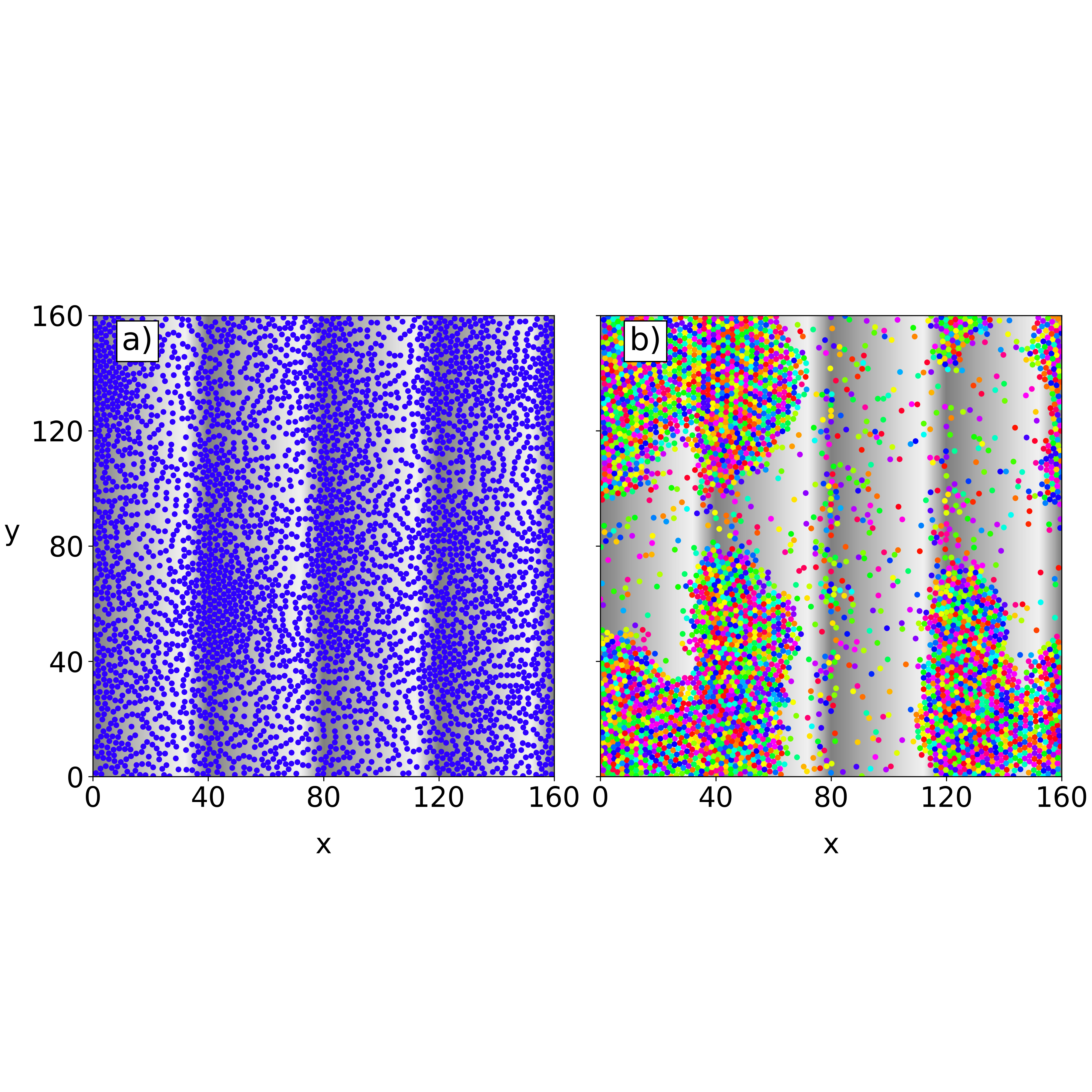}
\caption{Particle locations for the system in
Fig.~\ref{fig:10} at $r_r = 0.8$, $a_r = 40$, $f_r^{\rm max}=0.5$, and
$f_m=1.0$.
Color indicates the value of the phase $\phi_i$ for each particle.
(a) The labyrinth state with $J=-1.0$ and $K=0.1$.
(b) The normal MIPS state with $J=0.0$ and $K=0.0$.}
\label{fig:11}
\end{figure}

In Fig.~\ref{fig:11}(a) we illustrate the labyrinth state
for the system in Fig.~\ref{fig:10}
at $f_r^{\rm max} = 0.5$, where the ratchet efficiency is reduced
compared to the MIPS state. For comparison,
Fig.~\ref{fig:11}(b) shows the normal MIPS state at the same value
of $f_{r}^{\rm max} = 0.5$.
The labyrinth state appears as a modulated fluid
containing some local heterogeneities.
The MIPS state forms a large compact cluster that does not undergo
significant ratcheting; however,
the smaller fluid-like patches that are able to
align with the substrate along the $y$-direction
do undergo ratcheting motion.

\begin{figure}
\includegraphics[width=\columnwidth]{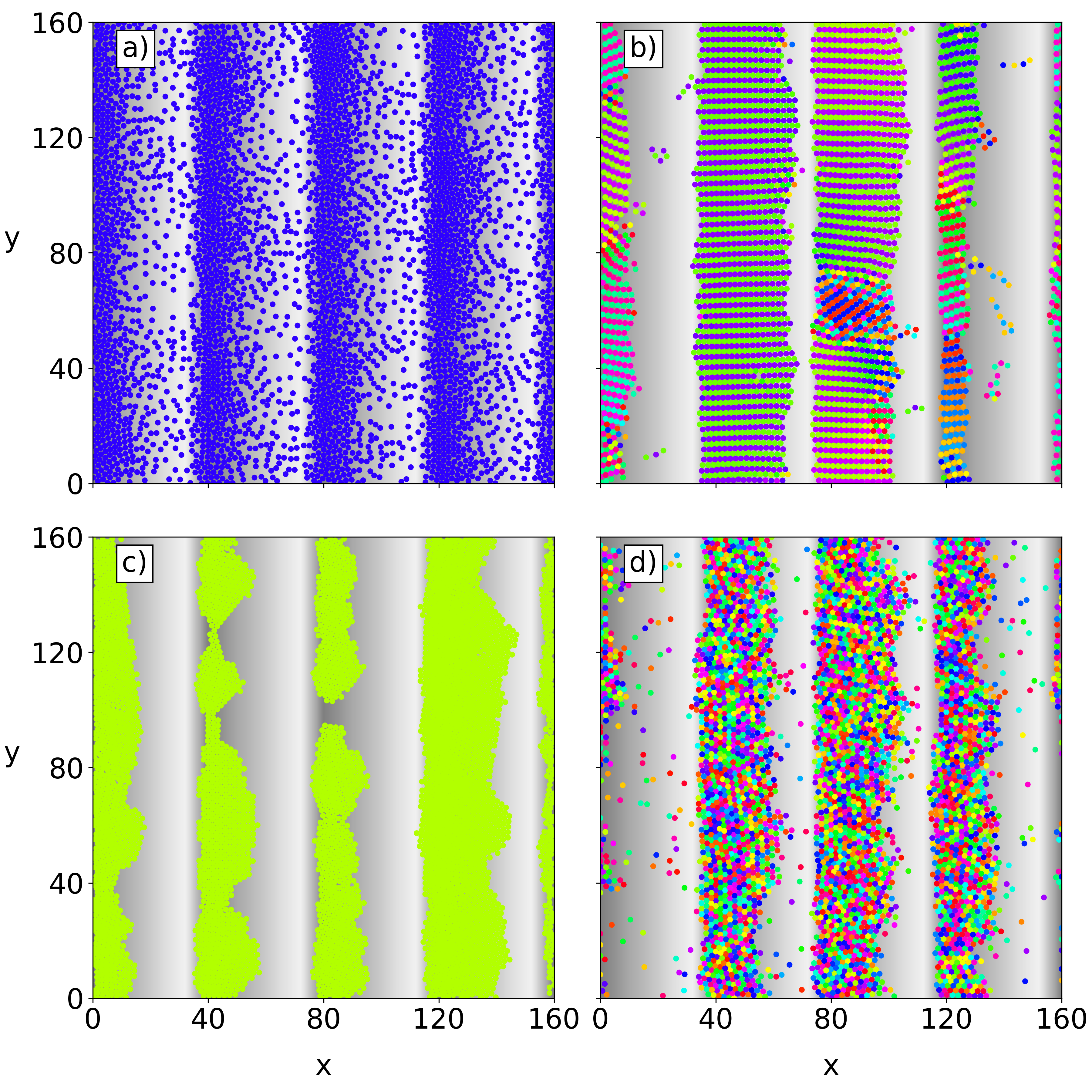}
\caption{Particle locations for the system
in Fig.~\ref{fig:10} at $r_r=0.8$, $a_r=40$,
$f_r^{\rm max} = 2.0$, and $f_m=1.0$.
Color indicates the value of the phase $\phi_j$ for
each particle.
(a) The labyrinth state with $J=-1.0$ and $K=0.1$ where
the strongest ratchet effect occurs.
(b) The frustrated MIPS state with $J=-1.5$ and $K=-0.1$.
(c) The active gel state with $J=1.0$ and $K=1.0$.
(d) The normal MIPS state with $J=0.0$ and $K=0.0$.}
\label{fig:12}
\end{figure}

Figure~\ref{fig:12}(a,b,c,d) shows the particles and
their individual phases at
$f_r^{\rm max} = 2.0$ for the labyrinth, frustrated MIPS,
active gel, and normal MIPS states from Fig.~\ref{fig:10}.
At this value of $f_r^{\rm max}$,
the labyrinth state exhibits the strongest ratcheting effect,
followed by the normal MIPS state, frustrated MIPS
state, and then the active gel.
The labyrinth state forms a well-defined density gradient inside each
substrate trough and includes
a number of quasi-one-dimensional particle chains aligned in
the $x$ direction.
At higher $f_r^{\rm max}$ values such as this one,
the different MIPS states form compact structures that can
align with the substrate in the $y$-direction.
An interesting feature of the frustrated MIPS state
is that stripe-like ordering appears in which
the phases of the individual particles are aligned in the $x$ direction.

\begin{figure}
\includegraphics[width=\columnwidth]{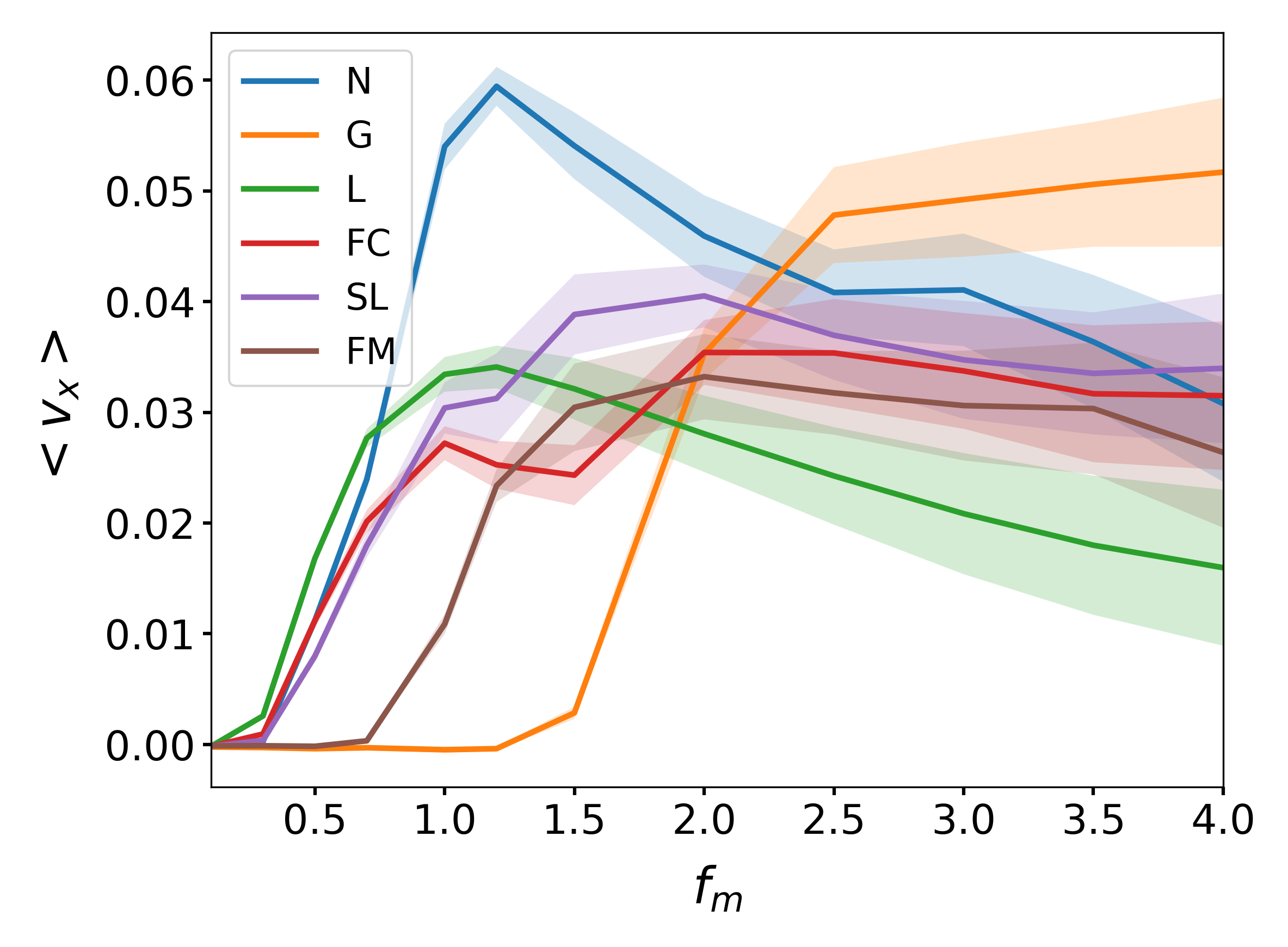}
\caption{$\langle V_x\rangle$ vs the motor force $f_m$
for a system with $r_r = 0.8$, $a_r=40$, and $f_r^{\rm max} = 1.0$
for the normal MIPS with $J=0$ and $K=0$ (N, blue),
active gel with $J=1.0$ and $K=1.0$ (G, orange),
labyrinth with $J=-1.0$ and $K=0.1$ (L, green),
frustrated cluster fluid with $J=1.5$ and $K=-0.1$ (FC, red),
superlattice MIPS with $J=1.0$ and $K=-0.1$ (SL, maroon),
and frustrated MIPS with $J=-1.5$ and $K=-0.1$ (FM, brown).
At $f_m = 0.0$ there is no ratchet effect,
while for large motor forces, the active gel shows the largest ratchet effect.}
\label{fig:13}
\end{figure}

We plot $\langle V_{x}\rangle$ vs the motor force $f_m$ at fixed
$r_r = 0.8$, $a_r = 40$, and $f_r^{\rm max} = 1.0$
for the six different states in Fig.~\ref{fig:13}.
At $f_m = 0.0$ there is no ratchet effect since activity is
needed for the active ratchet motion to occur.
The frustrated MIPS state does not exhibit a ratchet effect until
$f_m > 0.5$.
At $f_m = 1.0$, the normal MIPS
state shows a strong ratchet effect, while the active gel state
shows no ratchet effect.
For $f_m > 1.0$,
the active gel starts to exhibit a ratchet effect, and as $f_m$
increases, this ratcheting motion becomes greater than the ratcheting
in all of the other states.
At high $f_m$,
the amount of ratcheting motion in the normal MIPS, superlattice MIPS, and
frustrated cluster fluid is nearly equal.
The labyrinth state
has the strongest ratchet effect of all the states when $f_m < 0.6$.

\begin{figure}
\includegraphics[width=\columnwidth]{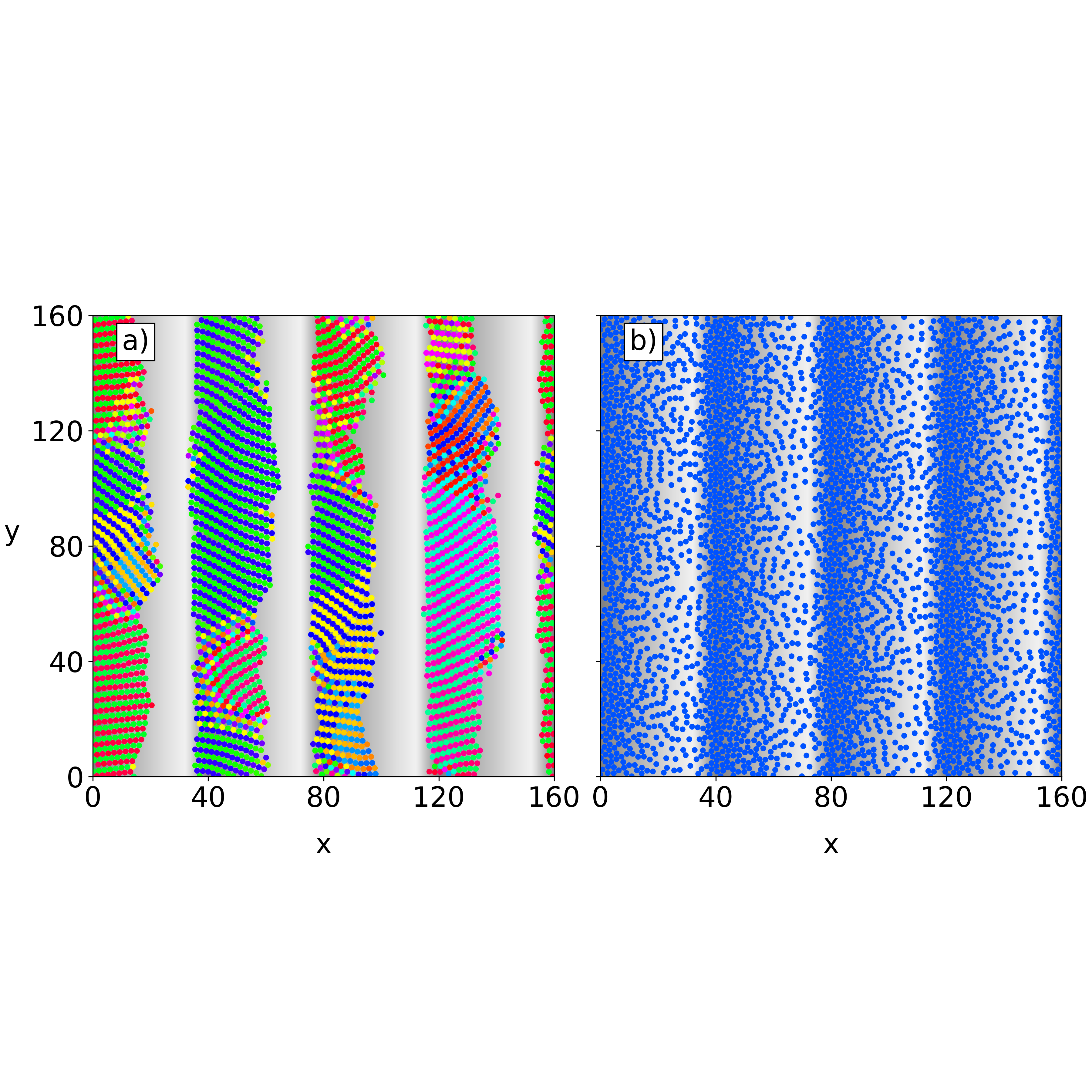}
\caption{Particle locations for the system in Fig.~\ref{fig:13}
at $r_r=0.8$, $a_r=40$, $f_r^{\rm max}=0.8$, and $f_m = 0.5$.
Color indicates the value of the phase $\phi_j$ for each particle.
(a) Superlattice MIPS state with $J=1.0$ and $K=-0.1$.
(b) Labyrinth state with $J=-1.0$ and $K=0.1$.}
\label{fig:14}
\end{figure}

In Fig.~\ref{fig:14}(a), we show the particle positions in the
superlattice MIPS state from Fig.~\ref{fig:13} at
$f_m = 0.5$ where the system forms dense
clusters aligned in the $y$-direction. This state has
no density gradient.
In contrast, for the same motor force,
the labyrinth phase illustrated in Fig.~\ref{fig:14}(b), has
a strong gradient in the particle density,
with a corresponding enhancement of the ratchet effect. 

\begin{figure}
\includegraphics[width=\columnwidth]{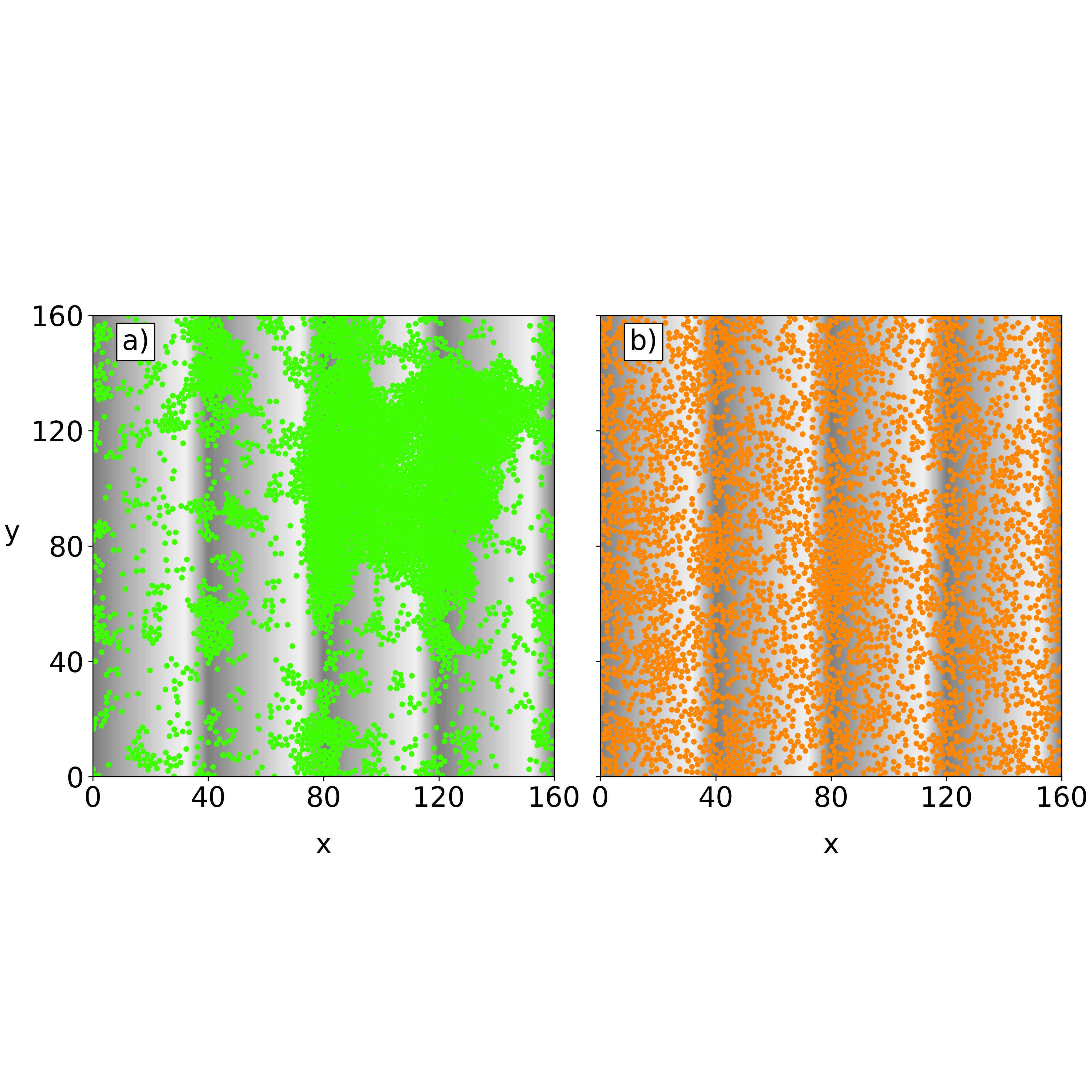}
\caption{Particle locations for the system in Fig.~\ref{fig:13} at
$r_r=0.8$, $a_r=40$, $f_r^{\rm max}=0.8$, and
$f_m = 3.5$.
(a) Active gel state at $J=1.0$ and $K=1.0$.
(b) Labyrinth state at $J=-1.0$ and $K=0.1$.}
\label{fig:15}
\end{figure}

At $f_m=3.5$ in the active gel state from Fig.~\ref{fig:13}, imaged 
in Fig.~\ref{fig:15}(a),
there is a strong ratchet effect.
For this motor force value the activity
is strong enough to partially break up the gel
structure, producing multiple groups of
smaller fast-moving clusters that can easily jump over the barrier
separating adjacent substrate minima.
For the same value of $f_m$, the labyrinth state in
Fig.~\ref{fig:15}(b) has a more uniform fluid
structure, and the density gradients that appeared for
lower values of $f_m$ are lost.
The MIPS states at this motor force
of $f_m=3.5$ still form dense clusters (not shown).

\begin{figure}
\includegraphics[width=\columnwidth]{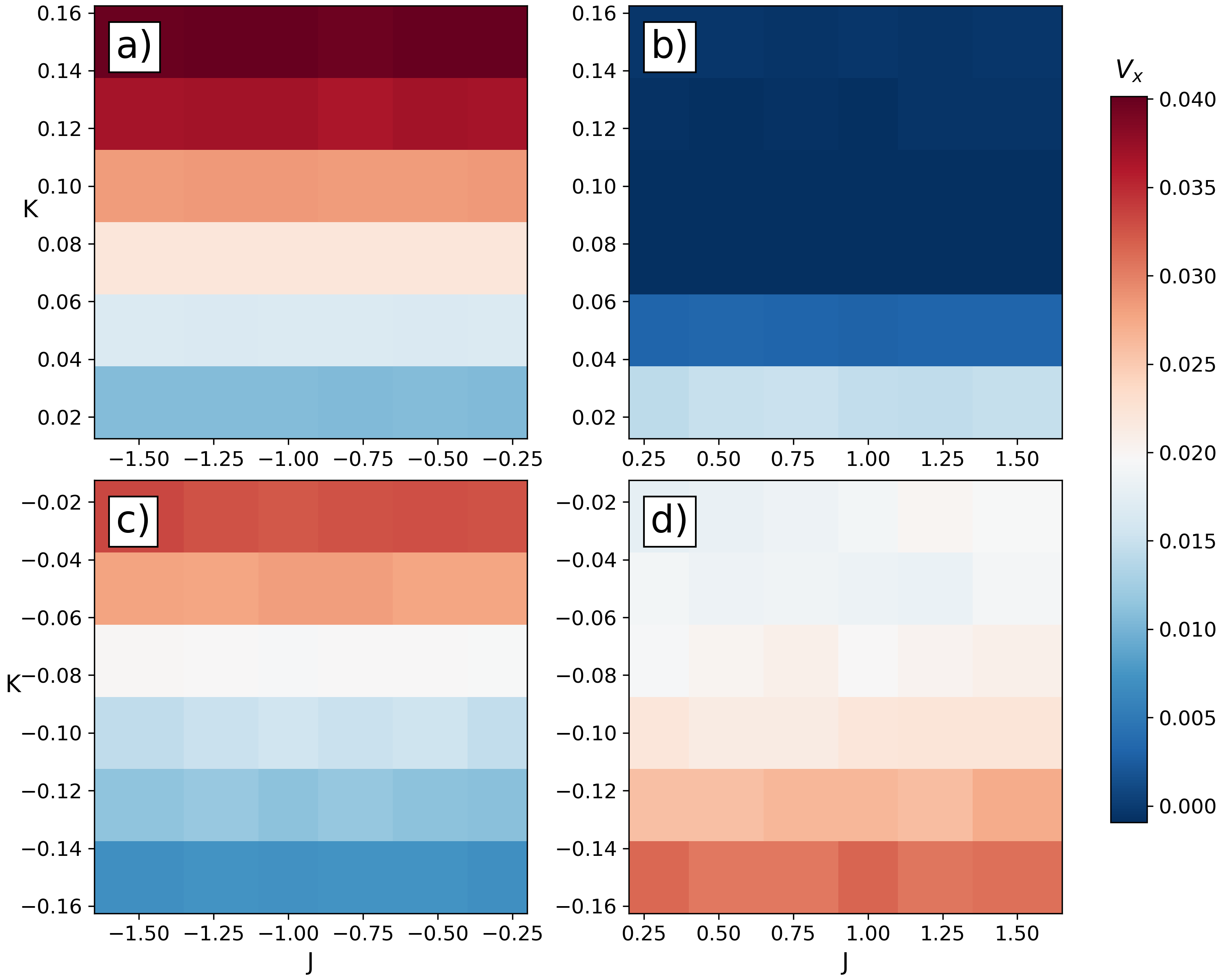}
\caption{Heat maps of $\langle V_x\rangle$ as a function of $K$ vs $J$
at $r_r = 0.8$, $a_r=40$, $f_r^{\rm max}=1.0$, and $f_m = 1.0$.
(a) The labyrinth regime with $K>0$ and $J<0$. 
(b) The active gel regime with $K>0$ and $J>0$.
(c) The frustrated MIPS regime with $K<0$ and $J<0$.
(d) The superlattice MIPS and frustrated cluster fluid
regime with $K<0$ and $J>0$.
}
\label{fig:16}
\end{figure}

In Fig.~\ref{fig:16}, we plot heat maps of the
ratchet velocity $\langle V_x \rangle$
as a function of $K$ versus $J$ for the different states in a system
with fixed $r_r = 0.8$, $a_r=40$, $f_r^{\rm max}=1.0$, and $f_m = 1.0$.
For $K > 0$ and $J > 0$ in Fig.~\ref{fig:16}(b),
the system forms an active gel state which shows the weakest
ratchet effect.
When $0 < K < 0.04$, the active gel has a small
finite ratchet effect.
In the labyrinth state, which appears for $K > 0.0$ and $J < 0.0$
in Fig.~\ref{fig:16}(a),
the amount of ratcheting increases with increasing $K$.
For $K < 0.0$ and $J < 0.0$ in Fig.~\ref{fig:16}(c),
the system is in the frustrated MIPS state. Here there is
a strong ratchet motion when the magnitude of $K$ is small, and
the ratchet effect weakens as the magnitude of $K$ increases.
In Fig.~\ref{fig:16}(d)
for $K < 0.0$ and $J > 0.0$,
the system is in the superlattice MIPS state
and frustrated cluster fluid state.
The ratchet efficiency increases as the magnitude of $K$ increases,
which is the opposite of the trend observed in Fig.~\ref{fig:16}(c)
for $K < 0.0$ and $J < 0.0$.
Ratcheting motion for the normal MIPS
state occurs along the $J = 0.0$ line, which is not shown in the figure.

\begin{figure}
\includegraphics[width=\columnwidth]{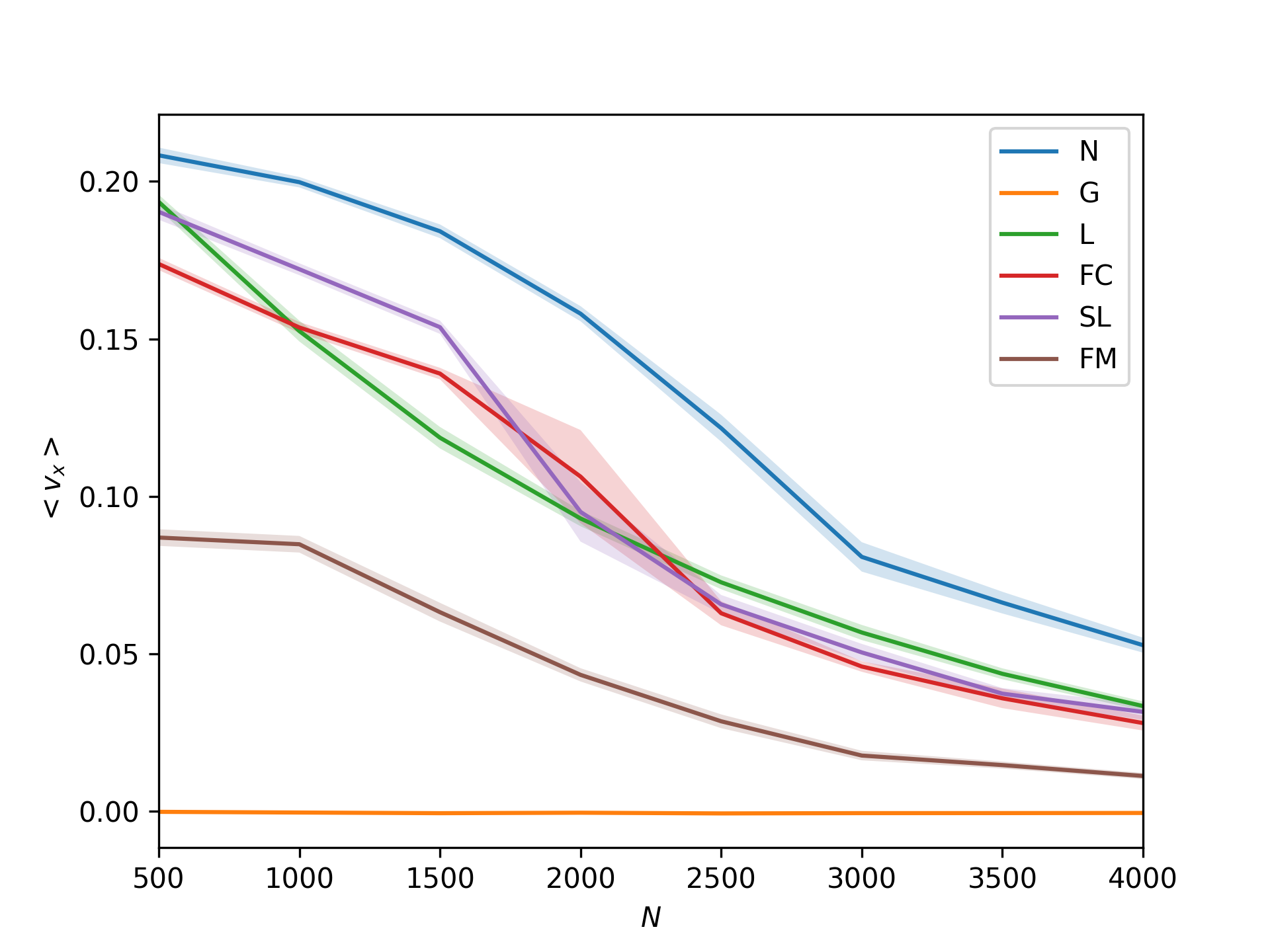}
\caption{$\langle V_x\rangle$ vs the number of particles $N$
for a system with $r_r = 0.8$, $a_r=40$, and $f_r^{\rm max} = 1.0$
for the normal MIPS with $J=0$ and $K=0$ (N, blue),
active gel with $J=1.0$ and $K=1.0$ (G, orange),
labyrinth with $J=-1.0$ and $K=0.1$ (L, green),
frustrated cluster fluid with $J=1.5$ and $K=-0.1$ (FC, red),
superlattice MIPS with $J=1.0$ and $K=-0.1$ (SL, maroon),
and frustrated MIPS with $J=-1.5$ and $K=-0.1$ (FM, brown).}
	\label{fig:17}
\end{figure}

In Figure ~\ref{fig:17}, we plot $\langle V_x\rangle$ versus the
number of particles $N$ in a system with $r_r=0.8$, $a_r=40$, and
$f_r^{\rm max}=1.0$.
In general, decreasing the particle number $N$
increases the amount of ratcheting
motion.
In the normal MIPS state,
the MIPS clusters disintegrate as $N$ is lowered from
$N=3000$ to $N=1500$.
Once the clusters are absent,
the individual particles
can ratchet more efficiently than the larger cluster objects.
In the superlattice MIPS and
frustrated cluster fluid states,
the clusters persist down to lower values of $N$
due to the additional attraction between particles from the swarmalator
interaction. As a result, the onset of the increase in ratchet motion
at lower particle densities shifts to smaller $N$.
In the frustrated MIPS state,
even when the main cluster disintegrates
as $N$ decreases, smaller
filamentary chains of particles with opposite phase form, which interferes
with the free ratcheting motion of individual particles and
lowers $\langle V_x\rangle$.
The gel state always retains
clusters even at $N=500$, and has no significant ratcheting motion
even for small particle densities.
The increase in ratcheting motion with decreasing $N$ is steepest for the
labyrinth state
since, in this state, the extra repulsion
from the swarmalator interaction improves the ability of
individual particles to ratchet.

\section{Summary}

We have examined the Adorj\'{a}rni swarmalator model in two
dimensions coupled to a quasi-one-dimensional asymmetric potential.
The Adorj\'{a}rni model consists of run-and-tumble active disks
that have an additional swarmalator force acting out to
twice the radius of the disk-disk interaction length.
This swarmalator interaction can be attractive or repulsive
and depends on the phase synchronization of the particles.
When there is no swarmalator force, we obtain
the standard run-and-tumble disk system that exhibits
motility-induced phase separation (MIPS).
When neighboring particles
preferentially synchronize to the same phase and the
swarmalator force is attractive, the system forms an active gel.
If the neighboring particles preferentially synchronize their phases
but the swarmalator force is repulsive, the system forms a labyrinth
state, and if the neighboring particles preferentially antisynchronize
their phases,
various other MIPS and fluid-like active cluster states appear.
We show that when this system is coupled to an asymmetric substrate,
the different states exhibit active matter ratcheting motion with
varying levels of efficiency depending on
the strength of the substrate, the substrate lattice spacing,
and the value of the active motor force.
When the motor force is approximately equal to
the maximum strength of the substrate,
the normal MIPS state has the most efficient ratcheting motion over
a range of parameters.
For small substrate lattice spacings, the labyrinth
state has the strongest ratchet effect since its quasi-one-dimensional
chains can align with the substrate troughs,
whereas the MIPS and cluster states only weakly couple to such
substrates, reducing the ratchet effect.
At larger substrate lattice spacings,
the different MIPS states can partially align with the substrate,
giving enhanced ratchet effects.
In general, the active gel state shows almost no ratchet effect;
however, when the motor force is large,
the active gel breaks apart and exhibits the strongest ratchet efficiency
compared to all of the other states.
Our results suggest that swarmalators or other synchronizing active
systems could exhibit a number of new
types of directed motion or active ratchet effects
beyond what is found for active systems with purely contact interactions.

\begin{acknowledgments}

This work was supported by the US Department of Energy through
the Los Alamos National Laboratory.  Los Alamos National Laboratory is
operated by Triad National Security, LLC, for the National Nuclear Security
Administration of the U. S. Department of Energy (Contract No. 892333218NCA000001).

\end{acknowledgments}

\bibliography{sw_ratchet}

\end{document}